\documentclass[aps,twocolumn,prx,nofootinbib,superscriptaddress,longbibliography]{revtex4-1}
\usepackage{amsmath}
\usepackage{amssymb}
\usepackage{amsfonts,txfonts}
\usepackage{bbold}
\usepackage{epsfig,color}
\usepackage{graphicx}
\usepackage{graphics}
\usepackage{dcolumn}
\usepackage{bm}
\usepackage{multirow}
\usepackage{bbm}
\usepackage{enumerate}
\usepackage{verbatim}
\usepackage{hyperref}
\usepackage{float}
\hypersetup{colorlinks=true,linkcolor=blue,anchorcolor=blue,citecolor=blue,filecolor=blue,urlcolor=blue,bookmarksnumbered=true,pdfview=FitB}
\usepackage[utf8]{inputenc}
\usepackage[english]{babel}
\usepackage[maxfloats=256]{morefloats}
\newcommand{\bn}{\boldsymbol{\nabla}}

\newcommand{\ve}{\varepsilon}

\newcommand{\bk}{{\bf k}}
\newcommand{\bp}{{\bf p}}

\newcommand{\bq}{{\bf q}}

\newcommand{\nn}{\nonumber}
\newcommand{\beq}{\begin{equation}}
\newcommand{\eeq}{\end{equation}}
\newcommand{\bea}{\begin{eqnarray}}
\newcommand{\eea}{\end{eqnarray}}
\newcommand{\bse}{\begin{subequations}}
\newcommand{\ese}{\end{subequations}}
\newcommand{\bwt}{\begin{widetext}}
\newcommand{\ewt}{\end{widetext}}
\newcommand\im{{\mathrm{Im}}}

\newcommand{\bv}{\boldsymbol{\varv}}
\newcommand{\bu}{{\bf u}}

\newcommand{\I}{\mathrm{Im}}
\newcommand{\R}{\mathrm{Re}}

\newcommand{\bsu}{\begin{subequations}}
\newcommand{\esu}{\end{subequations}}

\newcommand{\la}{\langle}
\newcommand{\ra}{\rangle}

\newcommand{\ofl}{\omega_{\text{FL}}}

\newcommand{\Eq}{Eq.~\eqref}

\newcommand{\EF}{E_{\text{F}}}

\newcommand{\bi}{\begin{itemize}}
\newcommand{\ei}{\end{itemize}}

\newcommand{\bj}{{\bf j}}
\newcommand{\cd}{c^\dagger}
\newcommand{\cnd}{c^{\phantom{\dagger}}}
\newcommand{\mB}{q_{\text{B}}}
\newcommand{\bnabla}{\boldsymbol{\nabla}}
\newcommand{\vf}{\varv_{\mathrm{F}}}
\newcommand{\kf}{k_{\mathrm{F}}}

\begin{document}
\title{Optical conductivity of a metal near an Ising-nematic quantum critical point}

\author{Songci Li} 
\thanks{Current address: Department of Physics and Tianjin Key Laboratory of Low Dimensional Materials Physics and Preparing Technology, Tianjin University, Tianjin 300354, China}
\affiliation{Department of Physics, University of Wisconsin-Madison, Madison, Wisconsin 53706, USA}

\author{Prachi Sharma} 
\affiliation{School of Physics and Astronomy, University of Minnesota, Minneapolis, Minnesota 55455, USA}

\author{Alex Levchenko} 
\affiliation{Department of Physics, University of Wisconsin-Madison, Madison, Wisconsin 53706, USA}

\author{Dmitrii L. Maslov}
\email{maslov@ufl.edu}
\affiliation{Department of Physics, University of Florida, Gainesville, Florida 32611, USA}

\date{October 31, 2023}

\begin{abstract}
We study the optical conductivity of a pristine two-dimensional electron system near an Ising-nematic quantum critical point. We discuss  
the relation between the frequency scaling of the conductivity and the shape of the Fermi surface, namely, whether it is isotropic, convex, or concave. We confirm the cancellation of the leading order terms in the optical conductivity for the cases of isotropic and convex Fermi surfaces and show that the remaining contribution scales as $|\omega|^{2/3}$ at $T=0$. On the contrary, the leading term, $\propto |\omega|^{-2/3}$, survives for a concave FS. We also address the frequency dependence of the optical conductivity near the convex-to-concave transition. Explicit calculations are carried out for the Fermi-liquid regime using the modified (but equivalent to the original) version of the Kubo formula, while the quantum-critical regime is accessed by employing the space-time scaling of the $Z=3$ critical theory.
\end{abstract}

\maketitle


\section{Introduction}\label{sec:Intro}

The optical conductivity of a correlated electron system contains important information about the strength of electron-electron (\emph{ee}) interaction. This information is encoded both in the renormalization of the Drude weight and in the frequency and/or temperature scaling of the conductivity. Experimental and theoretical studies of optical conductivity is a very active research area \cite{basov:2005,basov:2011,maslov:2017b,Armitage:2018,Tanner:book}.

Unlike the dc conductivity, which is rendered finite only either by umklapp \emph{ee} scattering \cite{Landau:1936, *Landau:collected,physkin} or Baber (electron-hole) scattering in compensated metals \cite{baber:1937,physkin}, the optical conductivity is relatively free of the constraints imposed by momentum conservation: as long as the electron spectrum is not parabolic, the optical conductivity is rendered finite by \emph{ee} interaction even in the absence of umklapps/compensation. In the Fermi liquid (FL) regime, the optical conductivity so far has been shown
to adhere to two scaling forms. The first one is due to Gurzhi \cite{gurzhi:1959}:
\bea
\sigma'(\omega,T)\propto \frac{\omega^2+4\pi^2 T^2}{\omega^2}.\label{Gurzhi}
\eea
This form occurs in the presence of umklapp scattering or compensation  both in two dimensions (2D) and three dimensions (3D), but also in 3D even without umklapps/compensation, as long as the electron dispersion contains higher than quadratic terms \cite{pal:2012b}. The second scaling form,
\bea
\sigma'(\omega,T)&\propto& \frac{\omega^2+4\pi^2 T^2}{\omega^2}(3\omega^2+8\pi^2 T^2)\nn\\
&&\times\left\{
\begin{array}{ccc}
\ln\max\{|\omega|,T\},\,\mathrm{2D}\\
1,\, \mathrm{3D}
\end{array}
\right.,\label{Dirac}
\eea
is believed to describe systems with isotropic but nonparabolic spectrum, e.g., 
Dirac metals \cite{Sharma:2021,Goyal:2023}, in the absence of umklapp scattering/compensation. \footnote{Although the primary focus of Ref.~\cite{Sharma:2021} is on Dirac metals, the analysis there is applicable to any nonparabolic spectrum.} At $T=0$, Eqs.~\eqref{Gurzhi} and \eqref{Dirac} are reduced to $\mathrm{const}$ and $\omega^2$, respectively (modulo a $\ln|\omega|$ factor in 2D). A suppression of $\sigma'(\omega,0)$ in Eq.~\eqref{Dirac} compared to Eq.~\eqref{Gurzhi} reflects partial Galilean invariance of an isotropic system. The limiting forms of Eq.~\eqref{Dirac}, i.e., $\omega^2$ and $T^4/\omega^2$, have been derived by a large number of authors in a variety of contexts \cite{gurzhi:1982,gurzhi:1987,gurzhi:1995, rosch:2005,rosch:2006,briskot:2015,levitov:2019}. Both scaling forms are valid for $\omega\gg 1/\tau_{\mathrm {j}}(T)$, where $\tau_{\mathrm{j}}(T)$ is the appropriate current relaxation time at finite $T$, which behaves as $1/T^2$ under the conditions when Eq.~\eqref{Gurzhi} is applicable or as $1/T^4$ (modulo a $\ln T$ factor in 2D) under the conditions when Eq.~\eqref{Dirac} is applicable.

The case of a 2D anisotropic Fermi surface (FS) (again, in the absence of umklapp scattering/compensation) is more complicated. There, the conductivity depends on whether the FS is convex or concave, i.e., on whether it has inflection points. In the dc case, the FL-like $T^2$ correction to the residual resistivity vanishes for a convex FS, while the surviving term behaves as $T^4\ln T$ \cite{gurzhi:1982, gurzhi:1987,gurzhi:1995,maslov:2011,pal:2012b}. On the other hand, the correction to the residual resistivity for a concave 2D FL is free of such cancellations and behaves as $T^2$. A similar cancellation for a convex FS  is expected to occur for the optical conductivity \cite{maslov:2017b} and, indeed, it has recently been shown to be the case for a non-FL at the Ising-nematic quantum critical point (QCP) \cite{Guo:2022}, while Ref.~\cite{Senthil:2022} arrived at the same result regardless of the shape (convex vs concave) of the FS.
The issues pertaining to the surviving term in the optical conductivity for a convex FS as well of the leading term for a concave FS have not been analyzed even in the FL regime. One of the goals of this paper is to resolve these outstanding issues. We will show explicitly that $\sigma'(\omega,T)$ for a 2D FL with convex/concave FS scales as predicted by Eqs.~\eqref{Dirac} and \eqref{Gurzhi}, respectively. 

Another goal of this paper is to generalize the results for an electron system at the Ising-nematic QCP, which is an example of the $D=2$, $Z=3$ criticality \cite{hertz:1976,millis:1993}. This subject has a long and somewhat controversial history. The optical conductivity of such a system [in the context of fermions coupled to a $U(1)$ gauge field] was claimed in Ref.~\cite{kim:1994} to behave as
\bea
\sigma'(\omega,0)\propto |\omega|^{-2/3}.\label{twothirds}
\eea
Later, it was realized, however, that the current-relaxing mechanism in Ref.~\cite{kim:1994}  was not properly specified \cite{maslov:2017b,chubukov:2017,Guo:2022,Senthil:2022,Senthil:2023}. As we already mentioned, subsequent studies demonstrated the vanishing of the leading $|\omega|^{-2/3}$ term for a convex FS \cite{Guo:2022}.
In this paper, we revisit the issue of the $-2/3$ scaling of the optical conductivity near an Ising-nematic QCP. We confirm that the $|\omega|^{-2/3}$ term vanishes both for an isotropic but nonparabolic spectrum and for a convex FS. We also show that the 
surviving term in the FL regime behaves according to Eq.~\eqref{Dirac} for both cases.\footnote{The actual form of the scaling function is somewhat different for the nematic system due to another contribution to the charge current, see Sec.~\ref{sec:ham}.} When extrapolated to the immediate vicinity of the Ising-nematic QCP, our result translates into 
\bea\label{QCP_np}
\sigma'(\omega,T)\propto |\omega|^{2/3}\max\left\{1,(T/|\omega|)^{8/3}\right\}.
\eea
The $T=0$ limit of our result, $\sigma'(\omega,0)\propto |\omega|^{2/3}$,  contradicts the $\sigma'(\omega,0)=\mathrm{const}$ result of Ref.~\cite{Guo:2022}. With respect to a concave FS, we show explicitly that the conductivity is restored back to the original form, Eq.~\eqref{twothirds}. For finite $T$, we find
\bea
\sigma'(\omega,T)\propto |\omega|^{-2/3}\max\{1, (T/|\omega|)^{4/3}\}.
\eea
The $T=0$ limit of this result was also obtained in Ref.~\cite{Senthil:2023} for a model of fermions coupled to loop current fluctuations. Following Refs.~\cite{pal:2012b} and \cite{pal:2012}, we also obtain the scaling form of the conductivity near the convex-to-concave transition. For the  reader's convenience, we summarize the results for the optical conductivity in various regimes in Table~\ref{table:Summary}.

\begin{table*}
    \begin{tabular}{|c|c|c|c|c|}\hline\hline
$\sigma'(\omega,T)$   & \multicolumn{4}{|c|}{Fermi surface}\\\hline
&
 \multicolumn{2}{|c|}{Isotropic with nonparabolic spectrum or convex}& \multicolumn{2}{|c|}{Concave}\\\hline
&  FL & QCP & FL &QCP\\
\hline
$\omega\gg T$ & $\omega^2\ln|\omega|$ & $|\omega|^{2/3}
$ & const &$|\omega|^{-2/3}$\\\hline
$\omega\ll T$ & $T^4\ln T/\omega^2$ & $T^{8/3}
/\omega^2$ & $T^2/\omega^2$ &$T^{4/3}/\omega^2$\\\hline\hline
    \end{tabular}
\caption{ Optical conductivity of a 2D electron system near an Ising-nemtaic quantum critical point (QCP) for different types of Fermi surfaces. FL stands for the Fermi-liquid region.\label{table:Summary}}    
\end{table*}

The rest of the paper is organized as follows. In Sec. \ref{sec:formalism}, we formulate the model and introduce the modified version of the Kubo formula that offers a practical advantage in calculation of the optical conductivity. In Sec. \ref{sec:isotropic}, we apply this formalism to an isotropic electron system with a nonparabolic energy spectrum. In Secs. \ref{sec:convex} and \ref{sec:concave}, we derive  
the optical conductivity of metals with convex and concave FSs, respectively, and also the near convex-to-concave transition. For each case, we discuss the frequency scaling of the conductivity both in the FL and quantum-critical regions.  In Sec. \ref{sec:concl} we summarize our main finding and provide a broader discussion. Appendices \ref{sec:app-A}--\ref{sec:app-C} present some technical details of our calculations.   


\section{General formalism}\label{sec:formalism}

\subsection{Hamiltonian and charge currents}\label{sec:ham}

We consider the following model Hamiltonian:
\bse
\begin{align}
&H=H_0+gH_{\text{int}},\label{Ham}\\
&H_0=\sum_{\bk s}\ve_\bk c^\dagger_{\bk,s} c^{\phantom{dagger}}_{\bk,s}\!\!\!\!\!,\label{Ham0}\\
&H_{\text{int}}=\frac 12\sum_\bq V(\bq)d_\bq d_{-\bq}\nn\\
&=\frac 12\sum_{\bk \bp \bq s s'}U(\bk,\bp,\bq)c^\dagger_{\bk_+,s}c^\dagger_{\bp_-,s'}
c^{\phantom{dagger}}_{\bp_+,s'}\!\!\!c^{\phantom{dagger}}_{\bk_-,s}\!\!\!,\label{Hint}
\end{align}
\ese
where $c^\dagger_{\bk,s} (c_{\bk,s})$ are the fermion creation (annihilation) operators for a particle in a state with momentum $\bk$ and spin projection $s$, 
\bea
\bk_\pm=\bk\pm\mathbf{q}/2,\;\mathrm{and}\; \bp_\pm=\bp\pm\mathbf{q}/2.\eea 
The coupling constant $g$ is factored out for convenience. Furthermore, $\ve_\bk=\epsilon_\bk-\EF$, $\epsilon_\bk$ is the band dispersion, $\EF$ is the Fermi energy, 
\bea
d_\bq=\sum_{\bk,s} F(\bk) c^\dagger_{\bk_+,s}c^{\phantom{dagger}}_{\bk_-,s}\label{dq}
\eea
is the charge density in  the angular momentum channel with a form-factor $F(\bk)$, and $U(\bk,\bp,\bq)=F(\bk)F(\bp)V(\bq)$. For simplicity, we assume that the FS has at least fourfold symmetry, although this assumption is by no means crucial. The form factor is normalized such that
\bea
  \int \frac{d\phi_\bk}{2\pi} F^2(\bk)=1,\label{norm}
  \eea
where $\phi_\bk$ is the azimuthal angle of $\bk$. For a purely density-density interaction, $F(\bk)=1$ and $U(\bk,\bp,\bq)=V(\bq)$. The interaction responsible for a quantum phase transition is modeled by the standard Orenstein-Zernike form
\bea\label{Vq}
V(\bq)=\frac{1}{q^2+\mB^2}, 
\eea
where $\mB$ is the bosonic mass, equal to the inverse correlation length of the order parameter fluctuations. 
We assume that the system is close to the QCP, such that $\mB\ll \kf$. At finite $\mB$, the system is in a FL regime at low enough energies
namely, for $\max\{\omega,T\}\ll \ofl$, where
\bea
\ofl=\vf \mB^3/gN_\mathrm{F},\label{ofl}
\eea
$\vf$ is the Fermi velocity, appropriately averaged over the FS, and $N_\mathrm{F}$ is the density of states at the Fermi energy.  In the FL regime, the dynamic interaction can be replaced by the static one, which is what we did in Eq.~\eqref{Vq}. Our explicit calculations will be carried out in the FL regime only. To extend the results to the immediate vicinity of the QCP, we will invoke the space-time scaling of the $Z=3$ critical theory and replace $\mB$ by $\max\{|\omega|^{1/3},T^{1/3}\}$ \cite{chubukov:2017}.

As usual, the (longitudinal part of) charge current is deduced from the continuity equation
\bea
\bq\cdot \bj_\bq=-[\rho_\bq,H],\label{cont}
\eea
where $\rho_\bq=e\sum_{\bk s} c^\dagger_{\bk_+,s} c^{\phantom{dagger}}_{\bk_-,s}$ is the charge density operator.
The commutator  $[\rho,H_0]$ 
yields the single-particle part of the current, the $\bq=0$ part of which is given by
\bea
\bj_{0}=e\sum_{\bk s} \bv_{\bk} c^\dagger_{\bk,s}c^{\phantom{dagger}}_{\bk,s}\!\!\!\!\!,\label{j0}
\eea
where $\bv_k=\boldsymbol\nabla_\bk\ve_\bk$ is the group velocity. However, the interaction part of the Hamiltonian Eq. \eqref{Hint} does not commute with the charge density and, therefore, there is one more contribution to the current.
Computing the commutator $[\rho_\bq,H_{\text{int}}]$,  we obtain for the interaction part of the current at $\bq=0$ (see Appendix \ref{sec:app-A} for details)
\bea
\bj_{\text{int}}=
e\sum_{\bk \bp \bq s s'}\left[
\left(\boldsymbol{\nabla}_{\bk} +\boldsymbol{\nabla}_\bp\right)U(\bk,\bp,\bq)\right]
c^\dagger_{\bk_+,s}c^\dagger_{\bp_-,s'}c^{\phantom{dagger}}_{\bp_+,s'}\!\!\!c^{\phantom{dagger}}_{\bk_-,s}\!\!\!\!\!,,\label{jint}
\eea
where $\bn_\bk=(\hat\phi/k)\partial_{\phi_\bk}$. The total current is given by the sum
\bea
\bj=\bj_0+g\bj_{\text{int}}.
\eea

\subsection{Modified Kubo formula for the optical conductivity}

Due to the fourfold symmetry of our model, $\sigma_{xx}=\sigma_{yy}$, and one can define the conductivity as $\sigma\equiv(\sigma_{xx}+\sigma_{yy})/2$. To calculate the conductivity, we will be using the modified version of the Kubo formula. The standard Kubo formula relates the real part of the conductivity to the imaginary part of the (retarded) current-current correlation function,
\bea
\sigma'(\omega,T)\equiv\R\,\sigma(\omega, T) =-\frac{1}{\omega}
\mathrm{Im}\,\Pi_\mathrm{j}(\omega,T),
\label{Kubo1}
\eea
where 
\begin{equation}
\Pi_\mathrm{j}(\omega,T)=
-\frac{i}{2} \int_{0}^\infty dt e^{i \omega t}
 \la [\bj(t)\stackrel{\cdot}{,}\bj(0)] \ra
 \equiv -\frac{i}{2}\la [\bj(t)\stackrel{\cdot}{,} \bj(0)]\ra_\omega
  \label{JJ}
\end{equation} 
where $[{\mathbf a};{\mathbf b}]={\mathbf a}\cdot {\mathbf b}-{\mathbf b}\cdot {\mathbf a}$. 
In this approach, the dissipative (real) part of the conductivity occurs only if the interaction is dynamic, which means that the bare static interaction in Eq.~\eqref{Hint} needs to be renormalized by dynamic particle-hole pairs.
An equivalent version of the Kubo formula is obtained integrating Eq.~\eqref{JJ} by parts \cite{rosch:2005,rosch:2006,Sharma:2021}
\bea
\sigma'(\omega, T) =\frac{1}{2\omega^3}\im\,\left\la
\left[{\bf K}(t)\stackrel{\cdot}{,}{\bf K}(0)\right]
\right\ra_\omega, \label{tJtJ}
\eea
where ${\bf K}(t)=i\partial_t \bj= [\bj,H]$. While Eq.~\eqref{Kubo1} contains a two-particle Green's function, Eq.~\eqref{tJtJ} contains a four-particle Green's function which, in general, is a more complicated object. However, to calculate the optical conductivity to order $\mathcal{O}(g^2)$, it suffices to  find ${\bf K}$ to order $\mathcal{O}(g)$:
\begin{equation}
{\bf K}(t)=[\bj_0+g\bj_{\text{int}},H_0+gH_{\text{int}}]=g
{\bf K}_1(t)+g{\bf K}_2(t)+\mathcal{O}(g^2),\label{K}
\end{equation}
where
\bse
\bea
{\bf K}_1(t)&=&[\bj_0,H_{\text{int}}],\label{jH1}\\
{\bf K}_2(t)&=&[\bj_{\text{int}},H_0].
\label{jH2}
\eea
\ese
Because the product of ${\bf K}(t)$ and ${\bf K}(0)$ in Eq.~\eqref{tJtJ} is already of order $g^2$, the averaging in this equation can be performed over the free-fermion states. The advantage of Eq.~\eqref{tJtJ} is that it accounts automatically only for current-relaxing processes without the need of combining diagrams for the current-current correlation function. In this approach, the renormalization of the bare static interaction by dynamic particle-hole pairs is accounted for automatically, when the product of two four-fermion correlators in Eq.~\eqref{tJtJ} is averaged over the 
quantum states of a noninteracting system.

Substituting Eq.~\eqref{K} into Eq.~\eqref{tJtJ}, we obtain the conductivity as the sum of three terms:
\bea
\sigma'(\omega,T)=\sigma'_1(\omega,T)+\sigma'_2(\omega,T)+
\sigma'_{12}(\omega,T),
\eea
where
\bse
\begin{align}
&\sigma'_1(\omega,T)=\frac{g^2}{2\omega^3}\im\,\la
\left[{\bf K}_1(t)\stackrel{\cdot}{,}{\bf K}_1(0)\right]\ra_\omega,\label{k1k1}\\
&\sigma'_2(\omega,T)=\frac{g^2}{2\omega^3}\im\,\la
\left[{\bf K}_2(t)\stackrel{\cdot}{,}{\bf K}_2(0)\right]\ra_\omega, \label{k2k2}\\
&\sigma'_{12}(\omega,T)=\frac{g^2}{2\omega^3}\im\left\{ \la
\left[{\bf K}_1(t)\stackrel{\cdot}{,}{\bf K}_2(0)\right]
\ra_\omega
- \la \left[{\bf K}_1(t)\stackrel{\cdot}{,}{\bf K}_2(0)\right]
\ra_{-\omega}\right\}\label{k1k2}.
\end{align}
\ese
For the commutators in Eqs.~\eqref{jH1} and \eqref{jH2} we obtain (see Appendix \ref{sec:app-B} for details)
\bse
\begin{align}
&{\bf  K}_1=-\frac{e}{2}\sum_{\bk\bp\bq s s'}\Delta\bv\,
U(\bk,\bp,\bq)c^\dagger_{\bk_+,s}c^\dagger_{\bp_-,s'} c^{\phantom{dagger}}_{\bp_+,s'}\!\!\!\!\!,c^{\phantom{dagger}}_{\bk_-,s}\label{K1} \\
 &{\bf K}_2=
 e\sum_{\bk \bp \bq s s'}
 \Delta\ve
 \left(\boldsymbol{\nabla}_{\bk}+\bn_\bp\right) U(\bk,\bp,\bq)c^\dagger_{\bk_+,s}c^\dagger_{\bp_-,s'} c^{\phantom{dagger}}_{\bp_+,s'}\!\!\!\!\!,c^{\phantom{dagger}}_{\bk_-,s}\label{K2}
\end{align}
\ese
where
\bea
\Delta\bv=\bv_{\bk_+}+\bv_{\bp_-}-\bv_{\bk_-}-\bv_{\bp_+}\label{Dv}
\eea
and
\bea
\Delta\ve=\ve_{\bk_+}+\ve_{\bp_-}-\ve_{\bk_-}-\ve_{\bp_+}\label{De},
\eea
are the changes in the total velocity and total energy of two fermions, respectively, due to a collision. The commutator in Eq.~\eqref{jH1} vanishes for a Galilean-invariant system, i.e., for $\bv_\bk=\bk/m$, and thus the conductivity for this case is not affected by the electron-electron interaction, as it should be.   

Substituting Eqs.~\eqref{K1} and \eqref{K2} into Eq.~\eqref{tJtJ} and applying Wick's theorem, we obtain for the three components of the conductivity
\bwt
\bse
\begin{align}
&\sigma'_1(\omega,T)=\pi e^2 g^2\frac{1-e^{-\omega/T}}{\omega^3}\int_{\bk,\bp,\bq} U^2(\bk,\bp,\bq)\left(\Delta\bv\right)^2M(\bk,\bp,\bq)
\delta(\ve_{\bk_+}+\ve_{\bp_-}-\ve_{\bk_-}-\ve_{\bp_+}+\omega),\label{sigma1}\\
&\sigma'_2(\omega,T)=4\pi e^2 g^2\frac{1-e^{-\omega/T}}{\omega^3}\int_{\bk,\bp,\bq}\left[ \left(\bn_\bk+\bn_\bp\right)U(\bk,\bp,\bq)\right]^2\left(\Delta\ve\right)^2M(\bk,\bp,\bq)
\delta(\ve_{\bk_+}+\ve_{\bp_-}-\ve_{\bk_-}-\ve_{\bp_+}+\omega),\label{sigma2}\\
&\sigma'_{12}(\omega,T)=4\pi e^2 g^2\frac{1-e^{-\omega/T}}{\omega^3}\int_{\bk,\bp,\bq} \Delta\ve \Delta\bv\cdot
\left(\bn_\bk+\bn_\bp\right)U^2(\bk,\bp,\bq)
M(\bk,\bp,\bq)\delta(\ve_{\bk_+}+\ve_{\bp_-}-\ve_{\bk_-}-\ve_{\bp_+}+\omega),\label{Sigma}
\end{align}
\ese
\ewt
where $\int_\bk$ is a shorthand for $\int d^2k/(2\pi)^2$, while 
\begin{equation}
M(\bk,\bp,\bq)=n_F(\ve_{\bk_+}) n_F(\ve_{\bp_-})[1-n_F(\ve_{\bk_-})][1- n_F(\ve_{\bp_+})],\label{M}
\end{equation}
with $n_F(\ve)$ being the Fermi function. In the equations above, we neglected the exchange part of the interaction, which is small compared to the direct part for a long-range interaction, considered in this paper.
 
The delta functions in Eqs.~\eqref{sigma2} and \eqref{Sigma} impose a constraint $\Delta\ve=-\omega$. Applying this constraint, we obtain instead of Eqs.~\eqref{sigma2} and \eqref{Sigma}
\bwt
\bse
\begin{align}
&\sigma'_2(\omega,T)=4\pi e^2 g^2\frac{1-e^{-\omega/T}}{\omega}\int_{\bk,\bp,\bq}\left[ \left(\bn_\bk+\bn_\bp\right)U(\bk,\bp,\bq)\right]^2
M(\bk,\bp,\bq)
\delta(\ve_{\bk_+}+\ve_{\bp_-}-\ve_{\bk_-}-\ve_{\bp_+}+\omega),\label{sigma2_b}\\
&\sigma'_{12}(\omega,T)=-4\pi e^2 g^2\frac{1-e^{-\omega/T}}{
\omega^2}\int_{\bk,\bp,\bq} \Delta\bv\cdot
\left(\bn_\bk+\bn_\bp\right)U^2(\bk,\bp,\bq)
M(\bk,\bp,\bq)\delta(\ve_{\bk_+}+\ve_{\bp_-}-\ve_{\bk_-}-\ve_{\bp_+}+\omega).\label{Sigma_b}
\end{align}
\ese
\ewt

Equations \eqref{sigma1}, \eqref{sigma2_b}, and \eqref{Sigma_b} form the basis of further analysis.
In what  follows, we will consider  three examples: the case of an isotropic but non-parabolic spectrum, as well as of convex and concave FSs. In all cases, we focus on a single-band system and neglect umklapp scattering. The reason for the last assumption is that a long-range interaction, characteristic for an Ising-nematic QCP, strongly suppresses umklapp scattering even if the FS occupies a substantial part of the Brillouin zone \cite{maslov:2011,pal:2012b}.
 
\section{Isotropic but nonparabolic spectrum}
\label{sec:isotropic}

In this section, we consider the case of an isotropic but nonparabolic spectrum, $\epsilon_\bk=\epsilon(k)$, which is encountered, e.g., in Dirac metals. We begin with $\sigma'_1(\omega,T)$ in Eq.~\eqref{sigma1}, the analysis of which
follows along the same lines as in Ref.~\cite{Sharma:2021}.
Relabeling the momenta as $\bk+\bq/2\to \bk$ and $\bp-\bq/2\to\bp$, introducing the energy transfer as $\Omega=\ve_{\bk-\bq}-\ve_{\bk}=\ve_{\bp}-\ve_{\bp
+\bq}+\omega$, and restricting the integrals over the fermionic momenta to narrow regions near the FS, we rewrite Eq.~\eqref{sigma1}
as 
\begin{widetext}
\begin{align}\label{sigma1-iso-np}
\sigma'_1(\omega,T)=&  
 e^2g^2\frac{\pi N_\mathrm{F}^2}{\omega^3} (1-e^{-\omega/T}) \int \frac{d^2q}{(2\pi)^2} 
 \int^{+\infty}_{-\infty} d \ve_\bk \int^{+\infty}_{-\infty} d\ve_\bp\int^{+\infty}_{-\infty} d\Omega 
 \int_0^{2\pi} \frac{d\phi_{\bk\bq}}{2\pi} \int_0^{2\pi} \frac{d\phi_{\bp\bq}}{2\pi} 
 F^2\left(\bk-\frac{\bq}{2}\right)F^2\left(\bp+\frac{\bq}{2}\right)V^2(\bq)\nn \\
&\times\left(\Delta \bv\right)^2   n_\mathrm{F}(\ve_{\bk}) n_\mathrm{F}(\ve_\bp) \left[1-n_\mathrm{F}
(\ve_{\bk}+\Omega)\right]\left[1-n_\mathrm{F}(\ve_\bp-\Omega+\omega)\right]
\delta(\Omega-\ve_{\bk-\bq}+\ve_{\bk} )
\delta(\Omega-\omega+\ve_{\bp+\bq}-\ve_{\bp}),
\end{align}
\end{widetext}
where $\phi_{{\bf n}{\bf m}}$ is the angle between vectors ${\bf n}$ and ${\bf m}$, and 
\bea
\Delta\bv=
\bv_\bk+\bv_\bp-\bv_{\bk-\bq}-\bv_{\bp+\bq}.
\label{Dv1}
\eea
Note that $\bv_{\mathbf{k}}=\epsilon^{\prime}(k)\mathbf{k}/k$ for an isotropic dispersion. Therefore, if we project all the momenta onto the FS, i.e., put  $k=p=|\mathbf{k}-\mathbf{q}|=|\bp+\bq|=k_F$ in $\Delta\bv$, then $\Delta\bv$ vanishes and so does the conductivity. To get a finite result, we need to expand $\Delta\bv$ around the FS. Performing such an expansion, we obtain
\bea
\label{vel}
\Delta\bv&=&\frac{w_{\text{np}}}{k_\mathrm{F}}\left[\left(\ve_\bk-\ve_{\bk-\bq}\right)\hat\bk+\left(\ve_\bp-\ve_{\bp+\bq}\right)\hat\bp\right.
\nn\\
&&\left.+\left(\ve_{\bk-\bq}-\ve_{\bp+\bq}\right)\frac{\bq}{k_\mathrm{F}}\right],
\eea
where $\hat\bk$ is the unit vector in the direction of $\bk$, etc.,
and 
\bea
w_{\text{np}}=1-\frac{k_F\epsilon''(k)}{\epsilon'(k)}\Big\vert_{k=k_F}
\eea
is the nonparabolicity coefficient \cite{Sharma:2021}.
A Galilean-invariant system has a parabolic spectrum , in which case $w_\mathrm{np}=0$. For a nonparabolic spectrum $w_\mathrm{np}\neq 0$; for example, $w_\mathrm{np}=1$ for the Dirac spectrum. 

Near quantum criticality, momentum transfers are small: $q\lesssim \mB\ll \kf$. Therefore,
 we can neglect the last term, proportional to $\bq$, in Eq.~\eqref{vel}, and also  
replace $F(\bk-{\bq}/{2})\approx F(\bk)$ and  $F(\bp+{\bq}/{2})\approx F(\bp)$ in Eq.~\eqref{sigma1-iso-np}.  Next, we express the dispersions entering Eq.~\eqref{vel}  via $\omega$ and $\Omega$, using the constraints imposed by the delta functions in Eq.~\eqref{sigma1-iso-np}. This yields
\bea
\left(\Delta \bv\right)^2=
\frac{w^2_{\text{np}}}{\kf^2}\left[
\Omega^2+(\Omega+\omega)^2-2\Omega(\Omega+\omega)\cos\phi_{\bk\bp} \right].\nn\\
\label{Dvsq}
\eea
It is to be expected (and will be shown below to be the case) that typical values of $|\Omega|$ are on  the order of $\max\{|\omega|,T\}$. Then $(\Delta \bv)^2\sim w^2_{\text{np}}\max\{\omega^2,T^2\}/\kf^2$, which implies that the conductivity is suppressed compared to its canonical FL value, Eq.~\eqref{Gurzhi}. To obtain the leading term in the conductivity, it suffices to neglect $\Omega$ and $\omega$ in the delta functions in Eq.~\eqref{sigma1-iso-np}. For small $q$, these delta functions are then reduced to the geometric constraints  $\cos\phi_{\bk\bq}=0$ and  $\cos\phi_{\bp\bq}=0$, which implies that $\bk$ and $\bp$ are either parallel or anti-parallel to each other, i.e., that $\cos\phi_{\bk\bp}=\pm 1$. Recalling the fourfold symmetry of the interaction,  we find that the contributions from $\cos\phi_{\bk\bp}=\pm 1$ in Eq.~\eqref{Dvsq} cancel each other, while the integral of the form factors over $\phi_\bq$ gives a constant
\bea
a=\int^{2\pi}_0 \frac{d\phi_\bq}{2\pi} F^4(\bq).
\eea
Next, the triple integral over $\ve_\bk$, $\ve_\bp$, and $\Omega$ is solved as \cite{Sharma:2021}
\begin{align}
&\iiint\limits^{+\infty}_{-\infty} d
\ve_\bk d\ve_\bp d\Omega \left[(\Omega+\omega)^2+\Omega^2\right]  \nn \\ 
&\times n_\mathrm{F}(\ve_\bk) n_\mathrm{F}(\ve_\bp) \left[1-n_\mathrm{F}(\ve_\bk+\Omega)\right]\left[1-n_\mathrm{F}(\ve_\bp-\Omega+\omega)\right]\nn\\ 
&=\frac{\omega^5}{
30(1-e^{-\omega/T})}\left(1+\frac{4\pi T^2}{\omega^2}\right)\left(3 +\frac{ 8 \pi^2 T^2}{\omega^2}\right).\label{energy_int}
\end{align}
Note that typical values of the variables in this integral are $|\ve_\bk|\sim |\ve_\bp|\sim |\Omega|\sim \max\{|\omega|,T\}$, which proves our earlier assertion. Finally, we notice that the integral over $q$ diverges logarithmically at the lower limit because each of the delta functions in Eq.~\eqref{sigma1-iso-np} brings in a factor of $1/q$. To regularize the divergence, one needs to return to the dynamic form of the interaction and ask how large $q$ should be for the Landau-damping term to be neglected. The answer is that $q\gg \mB |\omega|/\ofl$, where $\ofl$ is the upper boundary of the FL region \cite{Pimenov:2022} \footnote{We thank A. Chubukov for clarifying the cutoff procedure for us and for bringing Ref.~\cite{Pimenov:2022} to our attention.}, given by Eq.~\eqref{ofl}.  Therefore,
\bea
\int^\infty_{\mB|\omega|/\ofl} \frac{dq}{q} V^2(q)
\approx \frac{1}{\mB^4}\ln\frac{\ofl}{|\omega|}.
\eea

Collecting everything together, we arrive at
\bea
\sigma'_1(\omega,T)&=&\frac{e^2}{60\pi^2}a w^2_{\text{np}}
\frac{g^2N^2_{\mathrm{F}}}{\mB^4} \left(\frac{\omega}{\vf\kf}\right)^2\left(1+\frac{4\pi^2 T^2}{\omega^2}\right)\nn\\
&&\times\left(3 +\frac{ 8 \pi^2 T^2}{\omega^2}\right)\ln\left(\frac{\ofl}{\max\{\omega,T\}}\right),\label{sigma1r}
\eea
which is the scaling form advertised in Eq.~\eqref{Dirac}.

We now turn to $\sigma'_2(\omega,T)$ in Eq.~\eqref{sigma2_b}. 
The integrand of $\sigma'_2(\omega,T)$ can be projected right onto the FS without further expansions. The rest of the  calculation is the same as for $\sigma'_1(\omega,T)$, except for the energy integral, which yields
\begin{align}
\label{freqint2}
&\iiint\limits^{+\infty}_{-\infty} d
\ve_\bk d\ve_\bp d\Omega\, 
\left[1-n_\mathrm{F}(\ve_\bk+\Omega)\right]\left[1-n_\mathrm{F}(\ve_\bp-\Omega+\omega)\right]\nn \\ 
&\times  
n_\mathrm{F}(\ve_\bk) n_\mathrm{F}(\ve_\bp)=\frac{\omega^3}{
6(1-e^{-\omega/T})}\left(1+\frac{4\pi^2 T^2}{\omega^2}\right).
\end{align}
A straightforward calculation then leads to 
\bea
\sigma'_2(\omega,T)=\frac{e^2}{24\pi^2}\frac{g^2N^2_{\mathrm{F}}}{\mB^4} w_{\text{nem}}^2\left(\frac{\omega}{\vf\kf}\right)^2\left(1+\frac{4\pi^2 T^2}{\omega^2}\right)\nn\\ \times\ln\left(\frac{\ofl}{\max\{\omega,T\}}\right),
\label{sigma2r}
\eea
where
\bea
w^2_{\text{nem}}=\int^{2\pi}_0\frac{d\phi_\bq}{2\pi} \left(\frac{\partial F^2(\bq)}{\partial \phi_\bq}\right)^2
\eea
is the nematicity coefficient.

Finally, we come to the $\sigma'_{12}(\omega,T)$ contribution to the conductivity, Eq.~\eqref{Sigma_b}. By power counting, it is of the same order as $\sigma'_1$ and $\sigma'_2$ because $\Delta\varv\Delta\ve$ 
scales as $\omega^2$ [cf. Eq. (26)]. However $\sigma'_{12}(\omega,T)$ vanishes within the approximation of this section. The reason is that, once the last term in Eq.~\eqref{vel} is neglected, $\Delta\bv$  becomes a sum of two radial vectors, directed along $\hat\bk$ and $\hat\bp$, respectively.
As we already know, the leading contribution to the integral comes from almost collinear momenta $\bk$ and $\bp$. Therefore, $\Delta\bv$ is also a radial vector, collinear  with $\bk$ (or $\bp$).  Now, in Eq.~\eqref{Sigma_b} $\Delta\bv$ is dotted into ${\bf w}\equiv (\bn_\bk+\bn_\bp)U^2(\bk,\bp,\bq)$, which is a vector sum of two tangential vectors, proportional to $\hat\phi_\bk$ and $\hat\phi_\bp$, respectively. Again, due to collinearity of $\bk$ and $\bp$, vector ${\bf w}$ is perpendicular to both. Thus $\Delta\bv\cdot{\bf w}=0$, and the leading term  in $\sigma'_{12}(\omega,T)$ vanishes.

Therefore, the final result for the conductivity is the sum of Eqs.~\eqref{sigma1r} and \eqref{sigma2r}: 
\begin{align}
 &\sigma'(\omega,T)=\frac{e^2}{12\pi^2}\frac{g^2N^2_{\mathrm{F}}}{\mB^4} \left(\frac{\omega}{\vf\kf}\right)^2\left(1+\frac{4\pi^2 T^2}{\omega^2}\right)\nonumber \\ 
 &\times\left[\frac{a w^2_\mathrm{np}}{5}\left(3+\frac{8\pi^2 T^2}{\omega^2}\right)+\frac{w_\mathrm{nem}^2}{2}\right]\ln\left(\frac{\ofl}{\max\{|\omega|,T\}}\right).\label{np_res}
 \end{align}
The nonparabolic and nematic contributions to the conductivity are of the same order for $\omega\gtrsim T$, while the nonparabolic contribution is the dominant one for $\omega\ll T$. Schematically,  
\bea
\sigma'(\omega,T)\propto \mB^{-4} \omega^2\max\left\{\ln\frac{\ofl}{|\omega|},\frac{T^4}{\omega^4
}\ln \frac{\ofl}{T}\right\}.\label{combFL}
\eea
As explained in Sec.~\ref{sec:ham}, a crossover to the QCP is achieved by replacing 
$\mB\to \max\{|\omega|^{1/3},T^{1/3}\}$. At the same time, $\ofl$ is replaced by $\max\{|\omega|,T\}$, and thus the logarithmic factors are replaced by constants of order one. As a result, Eq.~\eqref{combFL} is replaced by Eq. \eqref{QCP_np}. 

Next, we turn our attention to anisotropic FSs. 

\begin{figure}[t!]
    \includegraphics[width=1.0\linewidth]{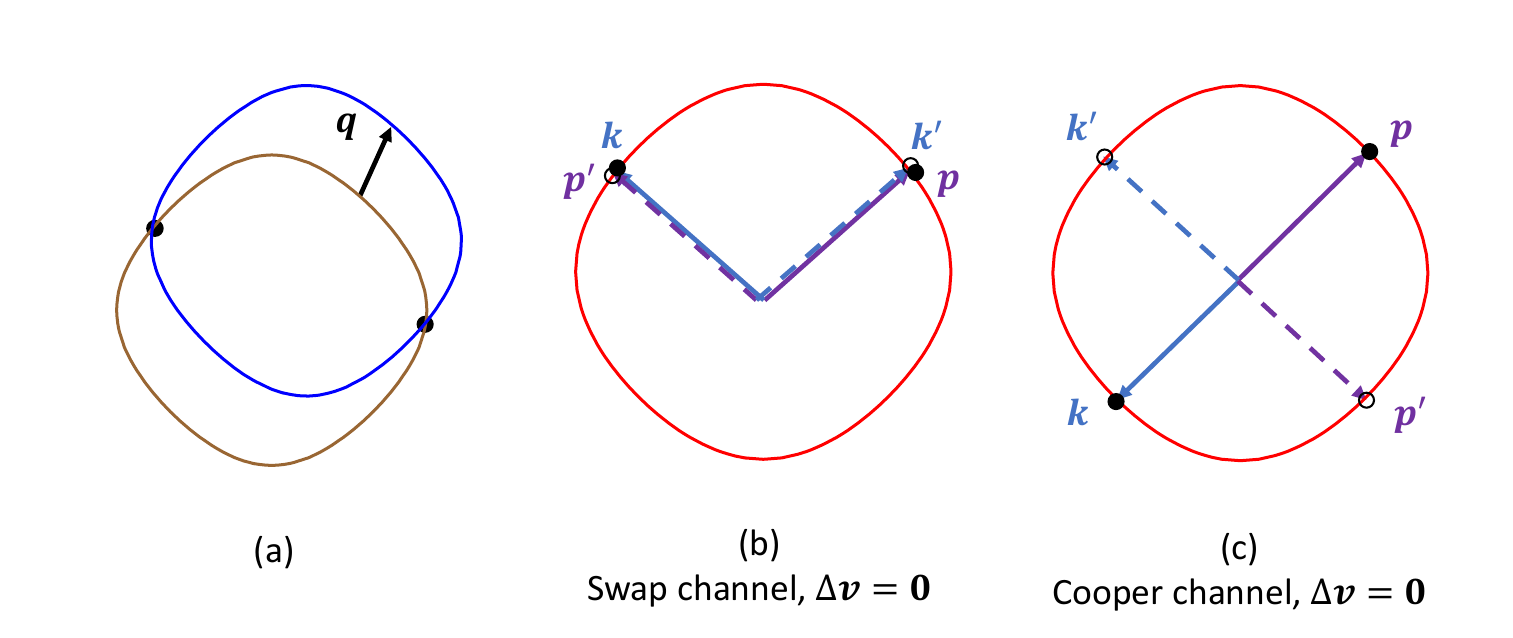}
    \caption{(a) Graphic solution of the equation $\ve_{\bk-\bq}=\ve_\bk$. As shown, a convex FS contour has no more than two self-intersection points (black dots).  (b) Swap scattering process, in which $\bk\to \bp'$  and $\bp\to \bk'$. (c) Cooper scattering process, in which $\bk+\bp=0=\bk'+\bp'$. Neither swap nor Cooper channel leads to current relaxation.}
    \label{fig:convex}
\end{figure}

\section{Convex Fermi surface}\label{sec:convex}

We proceed with the optical conductivity for a generic convex FS. As has been shown by many authors (see Sec.~\ref{sec:Intro}), the leading term in the conductivity, i.e., $\max\{\text{const},T^2/\omega^2\}$ in the FL regime, vanishes in this case. Our goal is to derive the surviving term.
Our main interest will be in the $\sigma'_1(\omega,T)$ contribution [Eq.~\eqref{sigma1}], because the general form of the $\sigma'_2(\omega,T)$ contribution [Eq.~\eqref{sigma2_b}] 
does not depend on the shape of the FS. 

We focus on the $T=0$ case first and begin with a brief review of the arguments for the vanishing of the leading term, as given in Refs.~\cite{pal:2012b} and \cite{maslov:2011}. To leading order in $\max\{\omega,T\}$, we project the integrand in Eq.~\eqref{sigma1-iso-np} onto the FS and neglect both frequencies in the delta functions ($\Omega$ and $\omega$). To ensure that the both delta functions are nonzero within the domain of integration, we need to find the solutions of two equations for the initial momenta $\bk$ and $\bp$ at fixed $\bq$:
\bse
\bea
\ve_{\bk-\bq}&=&\ve_\bk=0,\label{eqa}\\
\ve_{\bp+\bq}&=&\ve_\bp=0.\label{eqb}
\label{eqs}\eea
\ese
Excluding umklapps, the solutions of Eqs.~\eqref{eqa} and ~\eqref{eqb} are the points of intersection between the original FS contour and its two copied translated by $\pm \bq$, see Fig.~\ref{fig:convex}(a). Denoting $\bar\bp=-\bp$ and using that $\ve_{-\bk}=\ve_{\bk}$, we rewrite Eq~\eqref{eqb} as $\ve_{\bar\bp-\bq}=\ve_{\bar\bp}=0$, upon which it becomes identical to Eq.~\eqref{eqb}. 
For a convex FS, the first equation has (at most)  two solutions, see Fig.~\ref{fig:convex}(a). If we denote one of the solutions by $\bk_0$,  then the second one is $-\bk_0+\bq$. Indeed, $\ve_{-\bk_0+\bq-\bq}=\ve_{-\bk_0+\bq}\to \ve_{\bk_0}=\ve_{\bk_0-\bq}$. However, the equation for $\bar\bp$ also has (at most) two solutions: $(\bar\bp_0,-\bar\bp_0+\bq)=(-\bp_0,\bp_0+\bq)$. Because the equations are the same, the solutions must also coincide, which leaves us with two choices:  either $\bp_0=\bk_0-\bq$ or $\bp_0=-\bk_0$. The first choice corresponds to a ``swap'' scattering process, in which two electrons swap their initial momenta: $\bk_0\to \bk_0-\bq=\bp_0$, $\bp_0=\bk_0-\bq\to \bp_0+\bq=\bk_0$, see Fig.~\ref{fig:convex} (b). The second choice is a Cooper (or head-on) scattering process: $\bk_0\to \bk_0-\bq$, $\bp_0=-\bk_0\to \-\bk_0+\bq$; see Fig.~\ref{fig:convex}(c). According to Eqs.~\eqref{sigma1}-\eqref{Sigma}, a scattering process contributes to the conductivity only if $\Delta\bv=\bv_{\bk'}+\bv_{\bp'}-\bv_{\bk}-\bv_\bp\neq 0$. However, in a swap process $\bv_\bk'=\bv_\bp$ and $\bv_\bp'=\bv_\bk$, such that $\Delta\bv=0$. Likewise, in a Cooper process $\bv_{\bp}=\bv_{-\bk}=-\bv_\bk$ and $\bv_{\bp'}=\bv_{-\bk'}=-\bv_{\bk'}$, such that  $\Delta\bv=0$ again. Therefore, the leading term in the conductivity vanishes. 

To obtain a nonzero result, we have to consider small deviations from the previously found solutions for the swap  and Cooper channels, i.e., 
\bse
\bea
\bk&=&\bk_0+\delta\bk,\;\bp=\bp_0+\delta\bp=\bk_0-\bq+\delta\bp,\label{kp_sw}\\
\bk&=&\bk_0+\delta\bk,\;\bp=\bp_0+\delta\bp=-\bk_0+\delta\bp.\label{kp_C}
\eea
\ese
Expanding $\Delta\bv$  to first order in $\delta\bk$ and $\delta\bp$ around $\bk_0$ and $\bp_0$, we obtain
\bse
\bea
\Delta \bv_{\text{S}}&=&\left[(\delta\bk-\delta\bp)\cdot\bn\right](\bv_{\bk_0}-\bv_{\bk_0-\bq}),\label{swap}\\
\Delta \bv_{\text{C}}&=&\left[(\delta\bk+\delta\bp)\cdot\bn\right](\bv_{\bk_0-\bq}-\bv_{\bk_0})
\label{Cooper}
\eea
\ese
for the swap (S) and Cooper (C) channels, respectively.
Without a loss of generality, we choose $\omega>0$. With this choice and at $T=0$, the $\sigma'_1(\omega,0)$ contribution to the conductivity is reduced to  
\bwt
\bea
\sigma'_1(\omega,0)&=&\frac{\pi e^2 g^2}{(2\pi)^6\omega^3}\int d^2q \int d^2\delta k \int d^2\delta p \int^0_{-\omega} d\Omega\, F^2\left(\bk+\frac{\bq}{2}\right)F^2\left(\bp-\frac{\bq}{2}\right) V^2(\bq)(\Delta\bv)^2 \nn\\
&&\times\theta(\ve_\bk)\theta(-\ve_\bk-\Omega)\theta(\ve_\bp)\theta(-\ve_\bp+\Omega+\omega)\delta(\Omega-\ve_{\bk-\bq}+\ve_{\bk} )  \delta( \Omega+\omega+ \ve_{\bp+
\bq}-\ve_{\bp}).\label{T0}
\eea
\ewt
For any given $\bq$, we need to find solutions $\bk_0$ and $\bp_0$ of Eq.~\eqref{eqa} and \eqref{eqb}, integrate over $\delta \bk$ and $\delta \bp$ near these solutions, and then integrate over $\bq$. 

We start with the swap channel, in which  $\bp_0=\bk_0-\bq$. We expand the integrand of Eq.~\eqref{T0} in $\delta\bk$ and $\delta\bp$ and also use the condition $q\ll \langle\kf\rangle$, where $\langle\kf\rangle$ is 
a characteristic size of the FS. Then
\bwt
\bea
\sigma'_1(\omega,0)&=&\frac{\pi e^2 g^2}{(2\pi)^6\omega^3}\int d^2q F^4(\bk_0) V^2(\bq)\int d^2\delta k \int d^2\delta p \int^0_{-\omega} d\Omega\, (\Delta\bv_{\text{S}})^2\nn\\
&&\times
\theta(\bv_{\bk_0}\cdot\delta\bk)\theta(-\bv_{\bk_0}\cdot\delta\bk-\Omega)\theta(\bv_{\bk_0-\bq}\cdot\delta\bp)\theta(-\bv_{\bk_0-\bq}\cdot\delta\bp+\Omega+\omega)
\delta(\delta\bk\cdot\bu-\Omega)\delta(\delta\bp\cdot\bu-\Omega-\omega),\label{T01}
\eea
\ewt
where $\Delta\bv_\mathrm{S}$ is given by Eq.~\eqref{swap} and 
\bea 
\bu=\bv_{\bk_0-\bq}-\bv_{\bk_0}\approx -\left(\bq\cdot\bn\right)\bv_{\bk_0}.\label{u}
\eea
The delta functions in Eq.~\eqref{T01} imply that the 2D integrals over $\delta\bk$ and $\delta\bp$ are, in fact, one-dimensional integrals along the straight lines:
\bse
\bea
&&\delta\bk\cdot\bu=\Omega,\label{3}\\
&&\delta\bp\cdot\bu=\Omega+\omega.\label{4}
\eea
\ese
It is convenient to choose $\delta k_x$ and $\delta p_x$ as independent integration variables and exclude  $\delta k_y$ and $\delta p_y$ via
\bse
\bea
\delta k_y&=&\frac{\Omega}{u_y}-\delta k_x \frac{u_x}{u_y},\label{5}\\
\delta p_y&=&\frac{\Omega+\omega}{u_y}-\delta p_x \frac{u_x}{u_y}.\label{6}
\eea
\ese
The Pauli principle (imposed by the theta functions) and energy conservation (imposed by the delta-functions),
confine $\delta k_x$ and $\delta p_x$ to the following intervals:
\bse
\bea
k_{\min}\leq &\delta k_x&
\leq k_{\max},\label{kx}\\
p_{\min}\leq &\delta p_x&
\leq p_{\max},
\label{px}
\eea
\ese
where
\bse
\begin{align}
&k_{\min}=\theta\left(\frac{\nu}{u_y}\right)\frac{\varv_{\bk_0-\bq,y}}{\nu}\Omega+
\theta\left(-\frac{\nu}{u_y}\right)\frac{\varv_{\bk_0,y}}{\nu }\Omega,\label{kmin}\\
&k_{\max}=\theta\left(\frac{\nu}{u_y}\right)\frac{\varv_{\bk_0,y}}{\nu}\Omega+
\theta\left(-\frac{\nu}{u_y}\right)\frac{\varv_{\bk_0-\bq,y}}{\nu }\Omega,\label{kmax}\\
&p_{\min}=\theta\left(\frac{\nu}{u_y}\right)\frac{\varv_{\bk_0,y}}{\nu}(\Omega+\omega)+
\theta\left(-\frac{\nu}{u_y}\right)\frac{\varv_{\bk_0-\bq,y}}{\nu }(\Omega+\omega),\label{kmin}\\
&p_{\max}=\theta\left(\frac{\nu}{u_y}\right)\frac{\varv_{\bk_0-\bq,y}}{\nu}(\Omega+\omega)+
\theta\left(-\frac{\nu}{u_y}\right)\frac{\varv_{\bk_0,y}}{\nu }(\Omega+\omega),\label{pmax}
\end{align}
\ese
and
\bea
\nu&=&\varv_{\bk_0-\bq,x}\varv_{\bk_0,y}-\varv_{\bk_0-\bq,y}\varv_{\bk_0,x}\nn\\
&\approx& \varv_{\bk_0,x}\left(\bq\cdot\bn\right)\varv_{\bk_0,y}-\varv_{\bk_0,y}\left(\bq\cdot\bn\right)\varv_{\bk_0,x},\label{nu}
\eea
with $\bu$ defined by Eq.~\eqref{u}. The last step in Eq.~\eqref{nu} again accounts for the smallness of $q$.

Substituting Eqs.~\eqref{5} and \eqref{6} into Eq.~\eqref{swap}, we obtain for $(\Delta\bv)^2$ near the swap solution:
\bea
(\Delta\bv_{\text{S}})^2&=&\left\{\left[\delta k_x\left(\partial_{k_x}-\frac{u_x}{u_y}\partial_{k_y}\right)+\delta p_y\left(\frac{u_y}{u_x}\partial_{k_x}-\partial_{k_y}\right)\right.\right.\nn\\
&&\left.\left.-
\frac{\Omega+\omega}{u_x}\partial_{k_x}+\frac{\Omega}{u_y}\partial_{k_y}\right]\bu\right\}^2.\label{Deltav_sq}
\eea

With all the constraints having been resolved, Eq.~\eqref{T01} is reduced to
\begin{align}
\sigma'_1(\omega,0)=\frac{\pi e^2 g^2}{(2\pi)^6\omega^3}\int d^2q F^4(\bk_0) V^2(\bq)\frac{1}{u^2_y}\nn\\
\times\int^0_{-\omega} d\Omega\int^{k_{\max}}_{k_{\min}} d\delta k_x
\int^{p_{\max}}_{p_{\min}} d\delta p_x(\Delta\bv_{\text{S}})^2,
\label{T02}
\end{align}
The power counting of $\sigma'_1(\omega,0)$ is already obvious  at this point. Indeed, the integrals over $\delta k_x$ and $\delta p_x$ give a factor of $\omega$ each. Another factor of $\omega^2$ comes from $(\Delta\bv_{\text{S}})^2$. Finally, the integral over $\Omega$ gives one more factor of $\omega$. Therefore, $\sigma'_1(\omega,0)\propto \omega^{-3} \times \omega^2\times\omega^2\times \omega=\omega^2$.

To find the dependence of $\sigma'_1(\omega,0)$ on the bosonic mass ($\mB$), we need to analyze the behavior of the integrand in Eq.~\eqref{T02} at $q\to 0$. For simplicity, let's assume that $\nu/u_y>0$. Then $k_{\max}-k_{\min}=(v_{\bk_0,y}-v_{\bk_0-\bq,y}) \Omega/\nu$. The numerator in the last formula vanishes as $q$ at $q\to 0$ but, according to Eq.~\eqref{nu}, $\nu$ also vanishes as $q$ and, therefore, the range of integration over $\delta k_x$ remains finite at $q\to 0$. The same is true for $p_{\max}-p_{\min}$. Next, $(\Delta\bv)^2$ in Eq.~\eqref{Deltav_sq} contains only the ratios $u_i/u_j$ and $\partial_{k_i}u_j/u_l$ with $i,j,l=x,y$. While $u_i$ by itself vanishes as $q$ [cf. Eq.~\eqref{u}], the ratios quoted above remain finite and, therefore, $(\Delta\bv_{\text{S}})^2$ remains finite at $q\to 0$. Finally, the factor of $1/u_y^2$ in Eq.~\eqref{T02} diverges as $1/q^2$, and thus the integral over $q$ diverges logarithmically at $q\to 0$. This is exactly the same divergence that we encountered in Sec.~\ref{sec:isotropic}. Cutting the divergence off in the same way as before, we arrive at the final result:
\bea
\sigma'_1(\omega,0)\propto \mB^{-4} \omega^2 \ln\frac{\ofl}{|\omega|}.
\label{sigma_conv_res}
\eea
The frequency dependence of the conductivity is the same as for an isotropic but non-parabolic spectrum at $T=0$ [cf.~Eq.~\eqref{combFL}].

The contribution from the Cooper channel is analyzed along the same lines. It can be readily shown that the limits of integration over $\delta k_x$ and $\delta p_x$ for this case remain the same as for the swap channel [cf. Eqs.~(\ref{kmin}-\ref{pmax})].  The only change compared to the swap case is in the relative sign of $\delta\bk$ and $\delta \bp$ in Eq.~\eqref{Cooper}, which is irrelevant for scaling. Therefore, the combined contribution of the swap and Cooper channels is still given by Eq.~\eqref{sigma_conv_res}.

The finite-temperature case does not require a special analysis because, in the FL regime, the integrals over the energies and over momenta tangential to the FS factorize. Therefore, the energy integration gives the same result as in Eq.~\eqref{energy_int}. 

As we already said, the scaling form of the $\sigma'_2(\omega,0)$  contribution is insensitive to the shape of the FS, hence Eq.~\eqref{sigma2r} applies to a convex FS as well. Next, the $\sigma'_{12}(\omega,0)$ contribution already contains an extra factor of $\omega$ [cf. Eq.~\eqref{Sigma_b}], while another factor of $\omega$ comes from the expansion of $\Delta\bv$ around the swap and Cooper solutions. 
Therefore,  the total conductivity for a convex FS  can be written as
\begin{align}
 \sigma'(\omega,T)=&e^2\frac{g^2N^2_{\mathrm{F}}}{\mB^4} \left(\frac{\omega}{\vf\kf}\right)^2\left(1+\frac{4\pi^2 T^2}{\omega^2}\right)\nonumber \\ 
 &\times\left[\alpha\left(3+\frac{8\pi^2 T^2}{\omega^2}\right)+\beta\right]\ln\left(\frac{
 \ofl}{\max\{|\omega|,T\}}\right),\label{convex_res_T}
 \end{align}
where $\alpha$ and $\beta$ are the dimensionless coefficients which depend on the details of the convex FS. The overall form of the conductivity is the same as for an isotropic FS, cf. Eq.~\eqref{np_res}.
 
Extrapolating Eq.~\eqref{convex_res_T} to the vicinity of the QCP in the same way as in Sec.~\ref{sec:isotropic}, we arrive again at Eq.~\eqref{QCP_np}. At $T=0$, the conductivity scales as
\bea
\sigma'(\omega,0)\propto |\omega|^{2/3}.
\label{sigma_conv_qcp}
\eea
This result disagrees with that by Guo \textit{et al.} \cite{Guo:2022}, who argued that $\sigma'(\omega,0)=\text{const}$. 


\begin{figure}[t!]    
\includegraphics[width=1.0\linewidth]{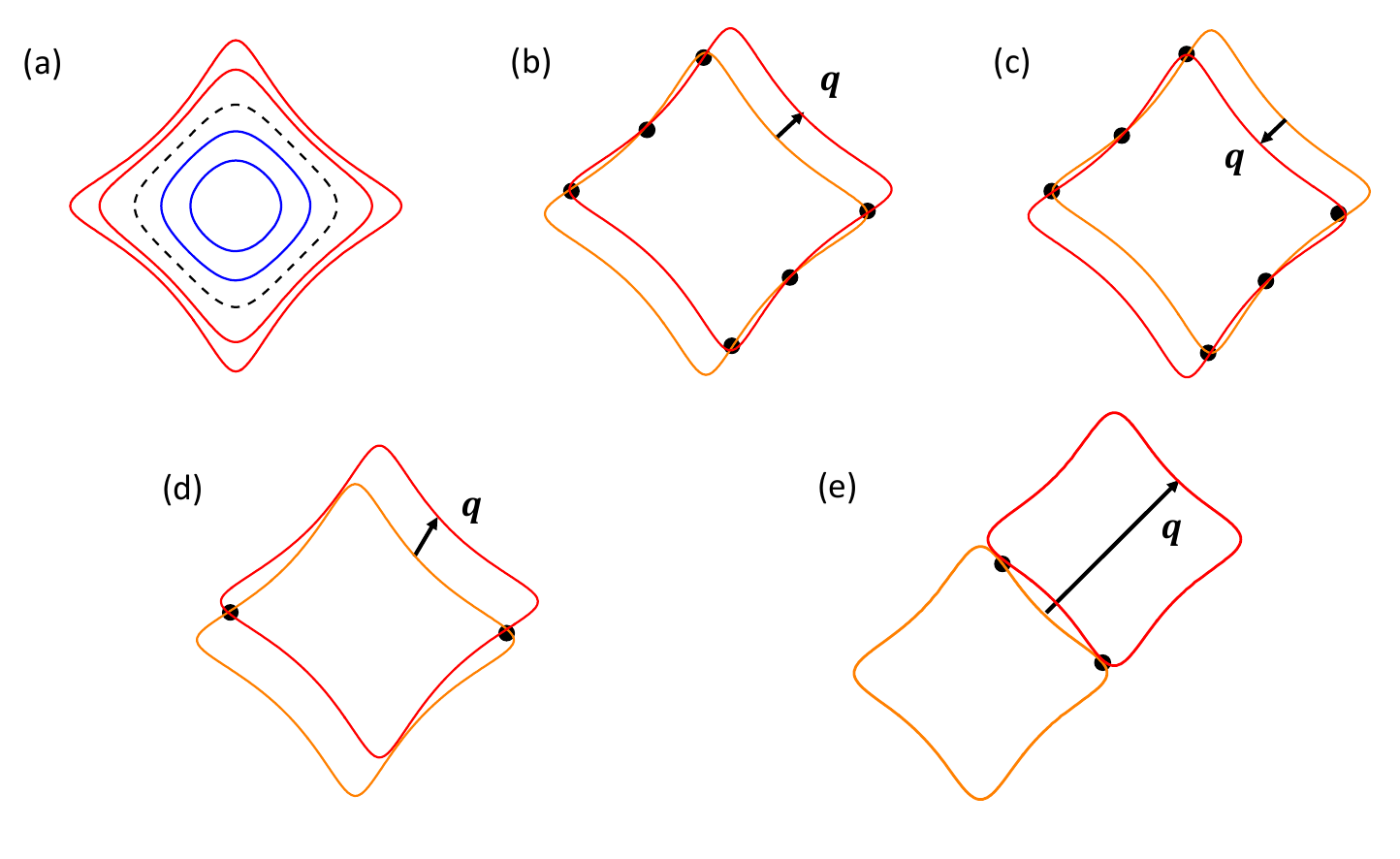}
    \caption{(a) Isoenergetic contours for a fourfold symmetric energy spectrum in 2D. As an example, we employ the tight-binding model with next-to-nearest neighbor hopping, for which
    $\ve(\bk)=-2t(\cos k_x+\cos k_y)+4rt\cos k_x \cos k_y$, and set $t=1$ and $r=0.3$. The convex-to-convex transitions occur at $E=E_c=8r(2r^2-1)$. The solid blue (red) curves are the convex (concave) Fermi surfaces. The dashed black curve corresponds to the critical value of Fermi energy at the convex-to-concave transition. (b) A concave FS contour can have more than two self-intersection points (six in the example shown). (c) Even for a concave FS, the number of self-intersection points can be smaller than the maximum number allowed if $\bq$ is pointed away from a special direction. (d) If the magnitude of the shift is sufficiently large, the number of self-intersection points is also less than the maximum number allowed.}
    \label{concave1}
\end{figure}

\begin{figure}[t!]  
\includegraphics[width=1.0\linewidth]{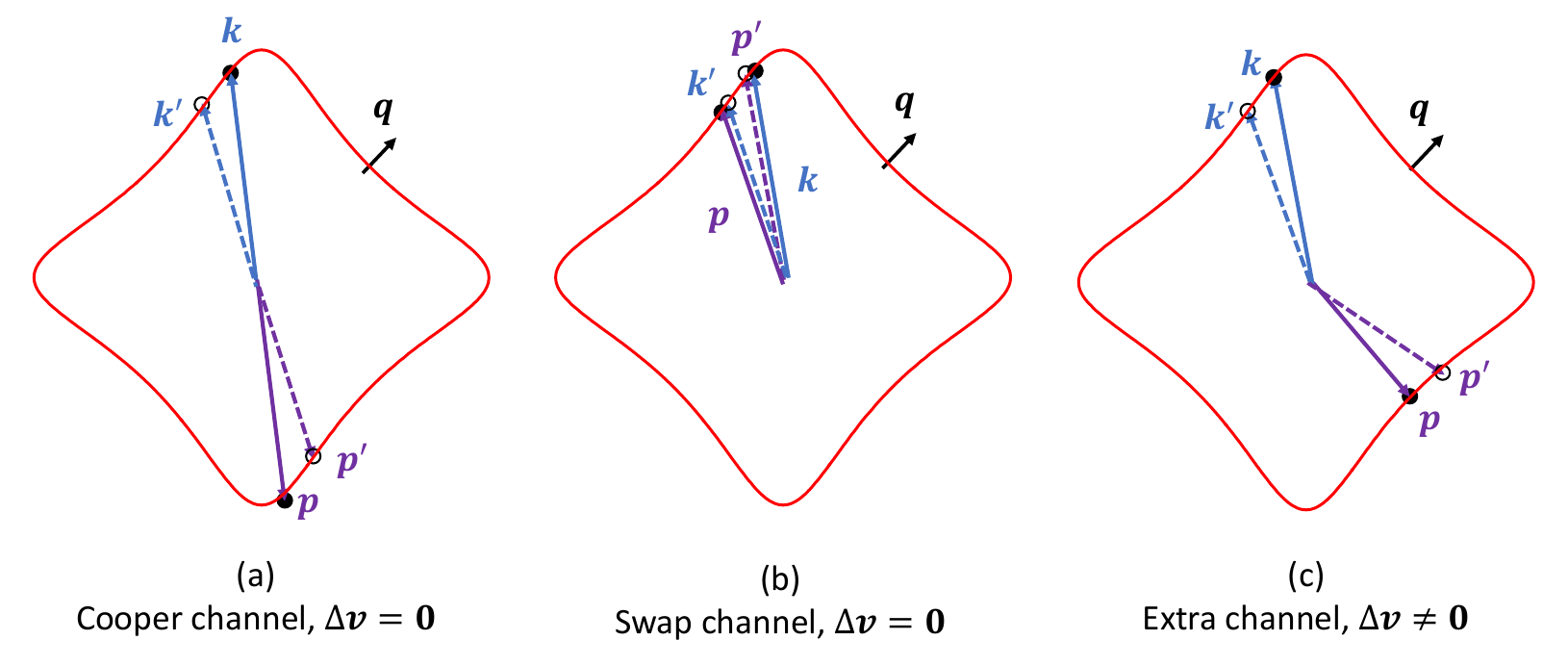}
    \caption{
    Scattering processes on a concave Fermi surface. (a) An example of Cooper 
   process with $\Delta \bv=0$. (b)  An example of swap 
   process, also with $\Delta \bv=0$. (c) A current-relaxing scattering 
   process with $\Delta \bv \neq 0$.}
    \label{concave2}
\end{figure}

\section{Concave Fermi surface}\label{sec:concave}
\subsection{Generic concave Fermi surface}\label{sec:concaveA}

A concave FS can have more than two self-intersection points. 
If the FS has mirror symmetry, the maximum number of such points is six, and it is achieved if the FS is translated along the mirror axis, as illustrated in Fig.~\ref{concave1}(b). One such point is the inflection point, two more are located on the same side of the curvilinear polygon, and the other three are located on the opposite side. Therefore, for given $\bq$, Eq.~\eqref{eqa} for the allowed initial momenta $\bk$ has up to six solutions. Solutions of Eq.~\eqref{eqb} for the initial momenta $\bp$ are obtained by translating the FS by $-\bq$, which yields another up to six solutions, see Fig.~\ref{concave1}(c).  The two sets of up to six solutions each generate up to 36 pairs of the initial momenta $\bk$ and $\bp$ at given $\bq$. These pairs contain not only Cooper [Fig.~\ref{concave2}(a)] and swap [Fig.~\ref{concave2}(b)] channels, but also scattering processes that do relax the current. An example of such process is shown in Fig.~\ref{concave2}(c). For any such process $\Delta\bv\neq 0$, and the conductivity in Eq.~\eqref{sigma1}-\eqref{Sigma} remains finite after projecting the integrands onto the FS.  Equation~\eqref{sigma1} for $\sigma'_1(\omega,T)$ is then reduced to
\bwt
\begin{align}
&\sigma'_1(\omega,T)=\frac{\pi e^2 g^2}{(2\pi)^4}\frac{1-e^{-\omega/T}}{\omega^3}\int^\infty_{-\infty} d\ve_\bk\int^\infty_{-\infty}  d\ve_\bp \int^\infty_{-\infty}  d\Omega\, 
n_\mathrm{F}(\ve_{\bk}+\Omega) n_\mathrm{F}(\ve_\bp - \omega-\Omega) \left[1-n_\mathrm{F}(\ve_{\bk})\right]\left[1-n_\mathrm{F}(\ve_\bp)\right]\nn\\
&\times
\int\frac{d^2q}{(2\pi)^2}\oint\frac{d\ell_\bk}{\varv_{\bk}}\oint\frac{d\ell_\bp}{\varv_{\bp}}\left(\Delta\bv\right)^2U^2(\bk,\bp,\bq)\delta\left(\ve_{\bk}-\ve_{\bk-\bq}+\Omega\right)
\delta\left(\ve_{\bp+\bq}-\ve_{\bp}+\omega+\Omega\right)\Big\vert_{\mathrm{FS}},\label{sigma1_concave} 
\end{align}
\ewt
where $d\ell_\bk$ is the line element of the FS contour, and it is understood that the fermionic momenta in the second line 
belong to the FS. As before, one can neglect the frequencies inside the delta-functions. In contrast to the convex case of Sec.~\ref{sec:convex}, $\Delta\bv$ does not contain additional dependence on the frequencies $\Omega$ and $\omega$ but, instead, it vanishes at $q\to 0$ as $q^2$. Therefore, the integral over $q$ is convergent at the lower limit (as opposed to being logarithmically divergent in the convex case) and gives $\int dq q q^2/q^2(q^2+\mB^2)^2\sim \mB^{-2}$.  Next, the triple integral over energies in the first line of Eq.~\eqref{sigma1_concave} is given by Eq.~\eqref{freqint2}. We thus obtain
\begin{align}
\sigma'_1(\omega,T)\propto  \mB^{-2} \frac{\omega^2+4\pi^2 T^2}{\omega^2}, \label{sigma1_concave_res}
\end{align}
which is just the Gurzhi scaling in Eq.~\eqref{Gurzhi}. 

Up to an overall factor, the $\sigma_2'$ contribution is still given by Eq.~\eqref{sigma2r}, which is subleading to Eq.~\eqref{sigma1_concave_res} and can be neglected.

The $\sigma'_{12}(\omega,T)$ contribution requires a bit of care. Rewriting \Eq{Sigma_b} in  the same way as \Eq{sigma1_concave}, we obtain
\bwt
 \bea
\label{sigma12_1M}
\sigma'_{12}(\omega,T) &=& - 
 \frac{e^2g^2}{4\pi^3} \frac{1-e^{-\omega/T}}{\omega^2} \int \frac{d^2q}{(2\pi)^2} V^2(\bq) \int^{+\infty}_{-\infty} d \ve_\bk \int^{+\infty}_{-\infty} d\ve_\bp\int^{+\infty}_{-\infty} d\Omega
 \oint \frac{d\ell_\bk}{\varv_\bk}
 \oint \frac{d\ell_\bp}{\varv_\bp}\nn\\
&&\times \left(\Delta \bv  \cdot  {\bf w}\right) n_\mathrm{F}(\ve_{\bk}
) n_\mathrm{F}(\ve_\bp 
) \left[1-n_\mathrm{F}(\ve_{\bk}+\Omega)\right]\left[1-n_\mathrm{F}(\ve_\bp-\Omega+\omega)\right]
\delta(\Omega-\ve_{\bk
-\bq}+\ve_{\bk} )  \delta( \Omega
-\omega+\ve_{\bp+\bq}-\ve_{\bp}),
\eea
\ewt
where 
\begin{equation}
{\bf w}=F^2(\bp+\bq/2)\bn_\bk F^2\vert_{\bk-\bq/2}+F^2(\bk-\bq/2)\bn_\bp F^2\vert_{\bp+\bq/2}.
\end{equation}
Suppose that we project all the momenta in the last equation onto the FS and set $T=0$. Then the triple integral over energies gives a factor of $\omega^3$ [cf. Eq.~\eqref{energy_int}], and thus $\sigma'_{12}(\omega,0)$ appears to scale as $\omega$. However, this is impossible because $\sigma'_{12}(\omega,0)$ must be even in $\omega$. Therefore, the leading, $\omega$ term must vanish identically, while the subleading term comes from expanding $\Delta\bv$ near the FS, as it was the case for an isotropic FS, cf. Sec.~\ref{sec:isotropic}. This implies that $\sigma'_{12}(\omega,0)$ scales at least as $\omega^2$ and is thus subleading to $\sigma_1'(\omega,0)$. An expanded 
discussion of this point is presented in Appendix \ref{sec:app-C}.

As before, a crossover to the QCP is achieved by replacing $\mB\to\max\{|\omega|^{1/3},T^{1/3}\}$ in Eq.~\eqref{sigma1_concave_res}, which yields
\bea
\sigma'(\omega,T)\propto \max\;\left\{\frac{1}{|\omega|^{2/3}},\frac{T^{4/3}}{\omega^2} \right\}.
\eea
This result restores the naive scaling form in Eq.~\eqref{twothirds}, obtained by substituting the current relaxation rate, $1/\tau_\mathrm{j}(\omega,T)\propto\max\{|\omega|^{4/3},T^{4/3}\}$, into the Drude formula $\sigma'(\omega,T)\propto 1/\omega^2\tau_{\mathrm{j}}(\omega,T)$.

\subsection{Optical conductivity near a convex-to-concave transition}\label{sec:concaveB}

Suppose that at a certain critical value of Fermi energy, $E_F=E_c$, 
the shape of the FS changes from convex ($\Delta\equiv E_F-E_c<0$) 
to concave ($\Delta>0$), see Fig.~\ref{concave1}(b). The results presented in Secs.~\ref{sec:convex} and \ref{sec:concaveA} are valid either well below or well above the convex-concave transition. Now, we consider the vicinity of the transition. 

Even if the FS is concave, the kinematic constraint $\ve_{\bk-\bq}=\ve_\bk$ has more than two solutions only if $\bq$ is along one of the high-symmetry axes and its magnitude is sufficiently small. If these conditions are not satisfied, then, as shown in Figs.~\ref{concave1} (d) and 2(e), the number of self-intersection points goes back to two and, again, only the Cooper and swap channels are possible. The width of the angular interval near a high-symmetry direction, $\delta\phi_\bq$, and the maximum value of $q,\, q_\mathrm{max}$, depend on $\Delta$ in a critical manner. In what follows, we will determine these parameters for the fourfold symmetric case and analyze the dependence of optical conductivity on $\omega$ and $\Delta$. In doing so, we will follow closely a derivation of the dc resistivity near the convex-to-concave transition
in the presence of disorder, presented in Ref.~\cite{pal:2012}. 

\begin{figure}    
\includegraphics[width=1.0\linewidth]{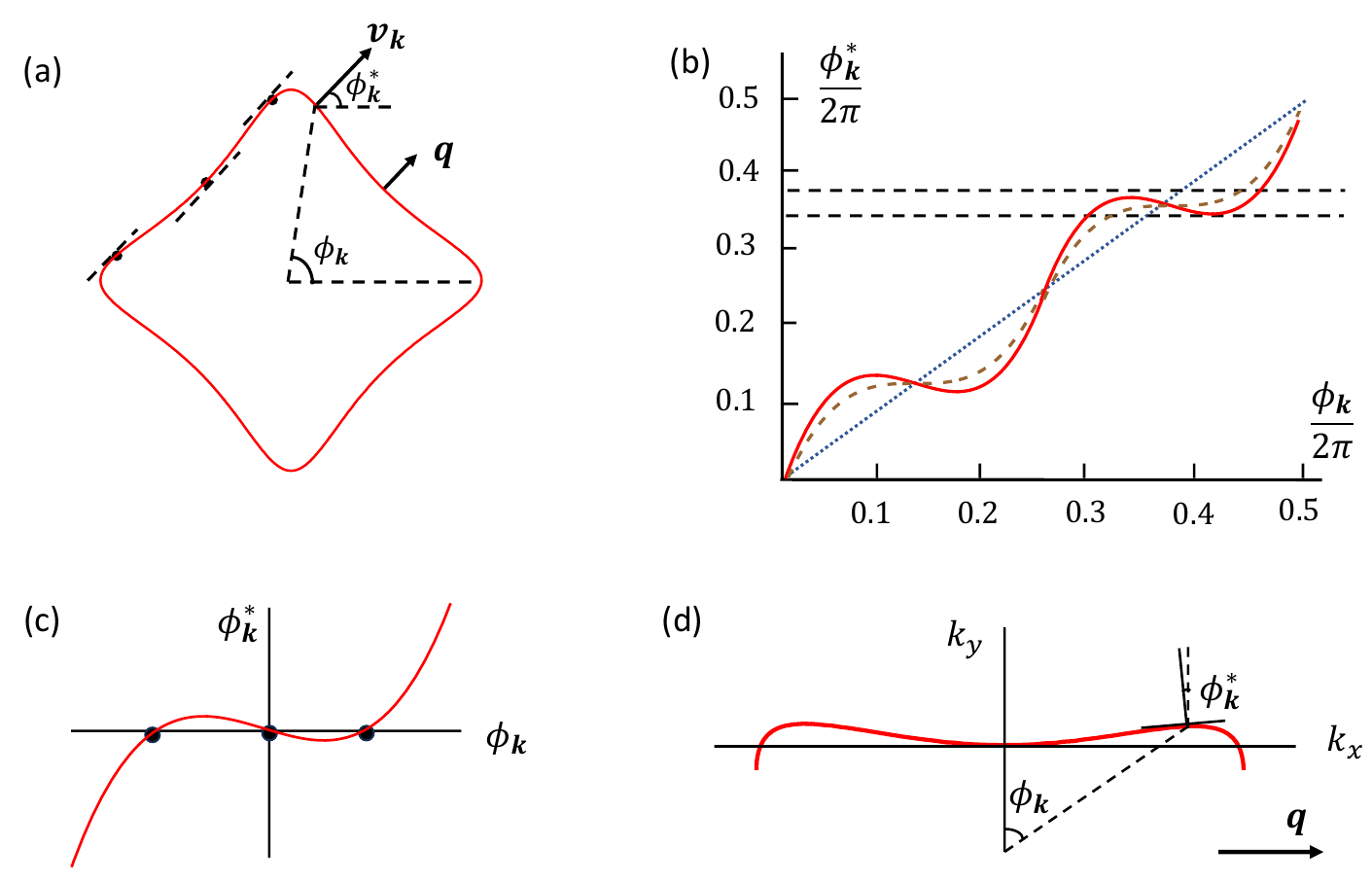}
    \caption{(a) At the self-intersection points (black dots), the momentum transfer $\bq$ is a tangent to the Fermi surface. $\bv_\bk$ is the Fermi velocity at point $\bk$ on the Fermi surface (FS). (b) The dependence of $\phi^*_\bk$ on $\phi_\bk$  for various types of the FS. Dashed curve: $E<E_c$, convex FS; 
    dotted line: at the convex-concave transition; solid 
    curve: $E>E_c$, concave FS. (c) An enlarged view of the non-monotonic dependence of $\phi^*_\bk$ on $\phi_\bk$. (d) A portion of the FS contour in local Cartesian coordinates.}
    \label{fig4}
\end{figure}

Let us first find $\delta\phi_\bq$. For $q\ll \kf$, the kinematic constraint $\ve_\bk=\ve_{\bk-\bq}$ is equivalent to $\bv_\bk \cdot \bq=0$. Since $\bv_\bk$ is along the normal to the FS at point $\bk$, this condition implies that the normals to the FS at the self-intersection points are perpendicular to $\bq$, see Fig.~\ref{fig4}(a). Let $\phi^*_\bk$ be an azimuthal angle of the normal to the FS at any given point $\bk$. Then the  constraint $\bv_\bk \cdot \bq=0$ can be written as $\phi^*_\bk=\phi_\bq+\pi/2$. (By symmetry it suffices to consider only the interval $\phi^*_\bk-\phi_\bq\in\{0,\pi\}$).  Given the relation between $\phi^*_\bk$ and $\phi_\bq$, one can find the corresponding angular interval $\delta\phi^*_\bk$ instead of $\delta\phi_\bq$. As one goes around the FS, $\phi^*_\bk$ varies with $\phi_\bk$. The difference between the convex and concave FSs is in that the dependence of $\phi^*_\bk$ on $\phi_\bk$ is monotonic for the former and non-monotonic for the latter, see 
Fig.~\ref{fig4}(b).  The equation $\phi^*_\bk=\phi_\bq+\pi/2$ has multiple roots only for a concave FS. Note that the curves $\phi^*_\bk(\phi_\bk)$ below and above the convex-to-concave transition coincide at inflection points. Near any of such points, the function $\phi^*_\bk(\phi_\bk)$  is described by a cubic polynomial [see Fig.~\ref{fig4}(c)], 
\bea\label{thetastar}
\phi^*_\bk(\phi_\bk)=b(\phi_\bk-\phi_{\mathrm{inf}})^3-c\Delta(\phi_\bk-\phi_{\mathrm{inf}}),
\eea 
where $\phi_{\mathrm{inf}}$ is the azimuthal coordinate of the inflection point, 
while $b$ and $c>0$ are constants that are specific for a given FS. The interval $\delta\phi^*_\bk$ is then equal to the vertical distance between the maximum and minimum of the curve, which gives
\begin{equation}
\delta\phi^*_\bk=\frac{
4c^{3/2}}{3\sqrt{3}
b^{1/2}}\Delta^{3/2}\sim \Delta^{3/2}.\label{deltaphi}
\end{equation}

Next, we need to find $\Delta \bv$.  For small $q$, we can write $\Delta \bv \approx (\bq\cdot\bnabla)\left(\bv_{\bk_1}-\bv_{\bk_2}\right)$, 
where $\bk_2$ and $\bk_1$ are any two solutions of the kinematic constraint $\bv_\bk\cdot\bq=0$. In terms of angles $\phi^*_\bk$ and $\phi_\bk$, we can rewrite  $(\Delta \bv)^2 \approx q^2\left(\frac{\partial\phi^*_\bk}{\partial\phi_\bk}\Big\vert_{\phi_{\bk_1}}
-\frac{\partial\phi^*_\bk}{\partial\phi_\bk}\Big\vert_{\phi_{\bk_2}}\right)^2
$; see Fig.~\ref{fig4}(d). For the $\phi^*_\bk(\phi_\bk)$ curve in Eq.~\eqref{thetastar}, we immediately find
\begin{equation}
(\Delta \bv)^2 
\propto (q\Delta)^2.\label{Deltav_concave}
\end{equation}

Finally, to find $q_\mathrm{max}$, we rewrite  Eq.~\eqref{thetastar} in a local Cartesian system with the $x$ axis along the tangent to the FS at the inflection point, see Fig.~\ref{fig4} (d). As shown in Ref.~\cite{pal:2012}, such an equation reads $k_y=-bk^4_x/12+c\Delta k^2_x/2$, where $k_{x,y}$ are measured from the inflection point \cite{pal:2012}. Using this relation to solve $\ve_\bk=\ve_{\bk-\bq}$, we obtain a cubic equation of $k_x$ which admits three different real roots only if $q\leq 2\sqrt{c \Delta/b}$. We thus arrive at following result:
\begin{equation}
q_\mathrm{max}\propto\Delta^{1/2}.\label{qmax}
\end{equation}

We now come back to the integral over $\bq$ in the second line of Eq.~\eqref{sigma1_concave}. Approximating $\int d^2q$ by $\int^{q_{\max}}_0 dqq\delta\phi_\bq\sim\int^{q_{\max}}_0 dqq\delta\phi^*_\bk$, using Eqs.~\eqref{deltaphi}--\eqref{qmax}, and substituting the interaction from Eq.~\eqref{Vq}, we obtain for the contribution to the conductivity from current-relaxing channels:
\begin{align}
\tilde\sigma'(\omega,0) \propto 
\delta\phi_\bk^* \Delta^2 \int^{q_\mathrm{max}}_0 \frac{dqq}{(q^2+\mB^2)^2}\propto \Delta^{7/2}\max\left\{\frac{1}{\mB^2},\frac{q^2_{\max}}{\mB^4}\right\}.
\end{align}
Adding up the contributions from the Cooper and swap channels, we obtain the conductivity in the FL regime, 
\bwt
\bea
\sigma'(\omega,0)\sim e^2g^2\left[\left(\frac{\Delta}{\EF}\right)^{7/2}\max\left\{\frac{\kf^2}{\mB^2},\frac{m\Delta \kf^2}{\mB^4}\right\}\Theta(\Delta)+\frac{\kf^4}{\mB^4}\left(\frac{\omega}{\EF}\right)^2\ln
\frac{\ofl}{|\omega|}\right],\label{FLctoc}
\eea
\ewt
where $\Theta(x)$ is the Heaviside function. The first term in the equation above is Gurzhi-like [cf. Eq.~\eqref{Gurzhi}], while the second one is the same as for a convex FS. 

Near the QCP, the first option for the Gurzhi-like term in Eq.~\eqref{FLctoc}, $\propto \Delta^{7/2}/\mB^2$, becomes $\Delta^{7/2}/|\omega|^{2/3}$. The second option,  $\propto \Delta^{9/2}/\mB^4$, formally becomes $ \Delta^{9/2}/|\omega|^{4/3}$ but it occurs only in the regime  where it is already subleading to the last term, and thus can be neglected. Finally, the optical conductivity near a convex-to-concave transition acquires the following scaling form:
\bea
&& \sigma'(\omega,0) \propto \Theta(\Delta)\frac{\Delta^{7/2}}{|\omega|^{2/3}}+|\omega|^{2/3}.
\eea 
Note that the conductivity has a minimum at $\omega=\omega_{\min}\propto \Delta^{21/8}$. 


\section{Summary}\label{sec:concl}

In this paper, we presented detailed calculations for the optical conductivity of a 2D metal near the Ising-nematic QCP. 
The focus of our attention was on the effect of the FS geometry that constrains the kinematics of 
possible scattering processes and thus ultimately determines the frequency dependence of the optical response.  We identified the swap and Cooper channels 
for the case of a convex FS, which do not relax current, and extra channels for the case of a concave FS, which do. On a technical level, we used the modified Kubo formula, which expresses the conductivity via the correlation function of the time derivatives of the current rather than that of the current itself \cite{rosch:2005,rosch:2006,Sharma:2021}. This method gives certain advantages as compared to a direct diagrammatic computation, namely it (i) bypasses the need to consider the dynamical interaction and (ii) automatically accounts for all current-relaxing processes. This obviates the need for the usual bookkeeping of various diagrams that tend to partially cancel each other for a generic non-Galilean--invariant FL. 

First, we considered  an electron system with a nonparabolic but isotropic dispersion relation. Our main result is given by Eq.~\eqref{np_res}, which generalizes prior related calculations \cite{Sharma:2021,Goyal:2023} by going beyond 
the model of a purely density-density interaction. Extrapolating Eq.~\eqref{np_res} to the quantum-critical regime of an Ising-nematic
phase transition, we find the frequency scaling of the optical conductivity as given by Eq. \eqref{QCP_np}.  

Next, we showed explicitly that Eqs.~\eqref{np_res} and \eqref{QCP_np} also hold for a generic convex FS. In doing so, we confirmed 
the vanishing of the leading term in the conductivity \cite{maslov:2017b,Guo:2022,Senthil:2022}. However, we disagree with the conclusion of Ref. \cite{Guo:2022} that the remaining term is independent of the frequency. 

Finally, we showed that the naive scaling form of the conductivity, $\sigma'(\omega)\propto 1/|\omega|^{2/3}$, is restored for a concave FS. In this respect, our results disagree with those of Ref.~\cite{Senthil:2022}, in which the conductivity was found to contain only the $\delta(\omega)$ term, regardless of the shape of the FS. We also derived the frequency dependence of conductivity at the onset of a convex-to-concave transition, see Eq. \eqref{FLctoc}.     

\textit{Note added in proof}: In a recent preprint arXiv:2311.03458 [cond-mat.str-el], H. Guo showed that the result of Ref.  \cite{Guo:2022} $\sigma'=$const for a convex Fermi surface is not valid.     


\acknowledgements

We thank A. Chubukov, L. Delacretaz, D. Else, I. Esterlis, Ya. Gindikin, H. Guo, I. Mandal, A. Patel, and S. Sachdev for fruitful discussions. This paper was supported by the National Science Foundation via Grants No. DMR-2203411 (S.L. and A.L.), No. DMR-2224000 (D.L.M.), and DMR-2011401 for MRSEC (P. S.). A.L. acknowledges hospitality of the Max Planck Institute for Solid State Research, where this work was performed in part, and a research fellowship funded by the Alexander von Humboldt Foundation. A.L. and D.L.M. acknowledge the hospitality of the Kavli Institute for Theoretical Physics (KITP), Santa Barbara, supported by the National Science Foundation under Grants No. NSF PHY-1748958 and No. PHY-2309135.

\begin{widetext}
\appendix

\section{Commutator algebra}\label{sec:app-A}

In this appendix, we will make use of the following two identities:
\bea
[A,BC]=[A,B]C+B[A,C]\label{id1}
\eea
and
\bea
[A,BC]=\{A,B\}C-B\{A,C\},\label{id2}
\eea
where $[x,y]$ and $\{x,y\}$ denote a commutator and anticommutator, respectively, of $x$ and $y$. To simplify notations, we will suppress spin indices of the fermionic operators, as they are not essential for the commutator algebra.
\label{app:Comm}
\subsection{Interaction part of the charge current}
\label{app:jint}
According to Eq.~\eqref{cont}, we need to calculate the following commutator:
\bea
\left[\rho_\bq,H_\mathrm{int}\right]=\sum_{\bq'} V(\bq') \left[\rho_\bq,d_{\bq'} d_{-\bq'}\right]
\eea
where $d_\bq$ is given by Eq.~\eqref{dq}. According to Eq.~\eqref{id1}:
\bea
\left[\rho_\bq,d_{\bq'} d_{-\bq'}\right]=[\rho_\bq,d_{\bq'}]d_{-\bq'}+d_{\bq'}[\rho_\bq,d_{-\bq'}].
\eea
Next, applying Eq.~\eqref{id1} and then Eq.~\eqref{id2}, we obtain
\bea
[\rho_\bq,d_{\bq'}]/e&=&\sum_{\bk,\bk'}F(\bk')\left[c^\dagger_{\bk+\bq/2}\cnd_{\bk-\bq/2},c^\dagger_{\bk'+\bq'/2}\cnd_{\bk'-\bq'/2}\right]\nn\\
&=&-\sum_{\bk,\bk'}F(\bk')\left(\left[\cd_{\bk'+\bq'/2},\cd_{\bk+\bq/2}\cnd_{\bk-\bq/2}\right]\cnd_{\bk'-\bq'/2}+\cd_{\bk'+\bq'/2}\left[\cnd_{\bk'-\bq'/2},\cd_{\bk+\bq/2}\cnd_{\bk-\bq/2}\right]\right)\nn\\
&=&-\sum_{\bk,\bk'}F(\bk')\left(\left(\left\{\cd_{\bk'+\bq'/2},\cd_{\bk+\bq/2}\right\}\cnd_{\bk-\bq/2}-\cd_{\bk+\bq/2}\left\{\cd_{\bk'+\bq'/2},\cnd_{\bk-\bq/2}\right\}\right)\cnd_{\bk'-\bq'/2}\right.\nn\\
&&\left.-\cd_{\bk'+\bq'/2}\left(\left\{\cnd_{\bk'+\bq/2},\cd_{\bk+\bq/2}\right\}\cnd_{\bk-\bq/2}-\cd_{\bk+\bq/2}\left\{\cnd_{\bk'-\bq'/2},\cnd_{\bk-\bq/2}\right\}\right)\right)\nn\\
&=&\sum_{\bk}F(\bk-\bq/2-\bq'/2)\cd_{\bk+\bq/2}\cnd_{\bk-\bq/2-\bq'}-F(\bk+\bq/2+\bq'/2)\cd_{\bk+\bq/2+\bq'}\cnd_{\bk-\bq/2}\nn\\
&=&\sum_{\bk}\left(F(\bk-\bq/2+\bq'/2)-F(\bk+\bq/2+\bq'/2)\right)\cd_{\bk+\bq/2+\bq'}\cnd_{\bk-\bq/2},
\eea
where we relabeled $\bk-\bq'\to\bk$  at the last step. For small $\bq$:
\bea
[\rho_\bq,d_{\bq'}]/e=-\sum_\bk \bq\cdot\bnabla_\bk F(\bk+\bq'/2)\cd_{\bk+\bq/2+\bq'}\cnd_{\bk-\bq/2}.
\eea
Likewise,
\bea
[\rho_\bq,d_{-\bq'}]=-\sum_\bk \bq\cdot\bnabla_\bk F(\bk-\bq'/2)\cd_{\bk+\bq/2-\bq'}\cnd_{\bk-\bq/2}.
\eea
The corresponding current operator is read from Eq.~\eqref{cont} as
\bea
\bj_\mathrm{int}(\bq)=e\sum_{\bk,\bp,\bq'}\left(\bnabla_\bk F(\bk+\bq'/2)F(\bp)\cd_{\bk+\bq/2+\bq'}\cd_{\bp-\bq'/2}\cnd_{\bp+\bq'/2}\cnd_{\bk-\bq/2}+\bnabla_\bp F(\bp-\bq'/2)F(\bk)\cd_{\bk+\bq'/2}\cd_{\bp+\bq/2-\bq'}\cnd_{\bp-\bq/2}\cnd_{\bk-\bq'/2}\right).
\eea
On putting $\bq=0$, relabeling $\bk+\bq'/2\to\bk$ in the first term and $\bp-\bq'/2\to\bp$ in the second one, and restoring spin indices, the last equation is reduced to Eq.~\eqref{jint} of the main text.

\subsection{Commutators of currents with the Hamiltonian}\label{app:commj}

We start with ${\bf K}_1=[\bj_0,H_\text{int}]$, where $H_0$ and $\bj_0$ are given by Eqs.~\eqref{Ham0} and \eqref{j0} of the main text, respectively. Applying Eq.~\eqref{id1}, we obtain
\bea
{\bf K}_1=\frac{1}{2}\sum_\bq V(\bq)\left([\bj_0,d_\bq]d_{-\bq}+d_\bq[\bj_0,d_{-\bq}]\right).
\eea
Evaluating the commutators in the above equation with the help of Eqs.~\eqref{id1} and \eqref{id2}, we find
\bea
[\bj_0,d_\bq]=e\sum_{\bk s} (\bv_{\bk+\bq/2}-\bv_{\bk-\bq/2})F(\bk)c^\dagger_{\bk+\bq/2} c^{\phantom{dagger}}_{\bk-\bq/2}\label{j0d}
\eea
and, correspondingly,
\bea
[\bj_0,d_{-\bq}]=e\sum_{\bp} (\bv_{\bp-\bq/2}-\bv_{\bp+\bq/2})F(\bp)c^\dagger_{\bp-\bq/2} c^{\phantom{dagger}}_{\bp+\bq/2}.
\eea
Combining the last two equations, we have
\bea
{\bf K}_1=-\frac{e}{2}\sum_{\bk \bp \bq}(\bv_{\bk+\bq/2}+\bv_{\bp-\bq/2}-\bv_{\bk-\bq/2}-\bv_{\bp+\bq/2}) U(\bk,\bp,\bq) c^\dagger_{\bk+\bq/2}  c^\dagger_{\bp-\bq/2} c^{\phantom{dagger}}_{\bp+\bq/2}c^{\phantom{dagger}}_{\bk-\bq/2}.
\label{K1A}
\eea
Restoring spin indices, we obtain Eq.~\eqref{K1} of the main text.

Next, we compute ${\bf K}_2=[\bj_{\text{int}},H_{0}]$, where $\bj_{\text{int}}$  is defined in Eq.~\eqref{jint} of the main text. With nematic density $d_\bq$, defined in Eq.~\eqref{dq},
and nematic current, defined  as
\bea
\boldsymbol{\mathcal{J}}_\bq=e\sum_\bk\boldsymbol{\nabla}_\bk F(\bk) c^\dagger_{\bk+\bq/2} c^{\phantom{dagger}}_{\bk-\bq/2},
\eea
the interacting part of the current $\bj_\text{int}$ can be re-written as
\bea
\bj_\text{int}=\frac{1}{2}\sum_\bq V(\bq) (\boldsymbol{\mathcal{J}}_\bq d_{-\bq}+\boldsymbol{\mathcal{J}}_{-\bq} d_\bq).
\eea
With the help of Eq.~\eqref{id1},  ${\bf K}_2$  becomes
\bea
{\bf K}_2&=&-\frac{1}{2}\sum_\bq V(\bq)\left([H_0,\boldsymbol{\mathcal{J}}_\bq]d_{-\bq}+\boldsymbol{\mathcal{J}}_\bq[H_0,d_{-\bq}]+[H_0,\boldsymbol{\mathcal{J}}_{-\bq}]d_\bq+\boldsymbol{\mathcal{J}}_{-\bq}[H_0,d_\bq]\right).\label{A15}
\eea
Calculating the commutators in the above equation, we obtain
\bea
[H_0,\boldsymbol{\mathcal{J}}_\bq]=e\sum_{\bk} \boldsymbol{\nabla}_\bk F(\bk) (\ve_{\bk+\bq/2}-\ve_{\bk-\bq/2}) c^\dagger_{\bk+\bq/2}c^{\phantom{dagger}}_{\bk-\bq/2}
\eea
and
\bea
[H_0,d_{\bq}]=\sum_\bp F(\bp) (\ve_{\bp+\bq/2}-\ve_{\bp-\bq/2}) c^\dagger_{\bp+\bq/2}c^{\phantom{dagger}}_{\bp-\bq/2}
\eea
Combining all the terms in Eq.~\eqref{A15}, we obtain
\bea
 {\bf K}_2=-e
 \sum_{\bk \bp \bq}(\boldsymbol{\nabla}_{\bk}+\boldsymbol{\nabla}_{\bp}) U(\bk,\bp,\bq)\left(\ve_{\bk-\bq/2}+\ve_{\bp+\bq/2}-\ve_{\bk+\bq/2}-\ve_{\bp-\bq/2}\right) c^\dagger_{\bk+\bq/2}  c^\dagger_{\bp-\bq/2}c^{\phantom{dagger}}_{\bp+\bq/2}c^{\phantom{dagger}}_{\bk-\bq/2}.
\eea
On restoring spin indices, the last result gives Eq.~\eqref{K2} of the main text.


\section{Averaging of commutators over the noninteracting system}\label{sec:app-B}

In this appendix, we derive Eqs.~\eqref{sigma1}--\eqref{Sigma}.
We begin with \Eq{k1k1}, which reads explicitly
\bea
\langle\left[{\bf{K}}_1(t),{\bf{K}}_1(0)\right]\rangle_{\omega} & = &-i\frac{e^{2}}{4}\int_{0}^{\infty}dte^{i\omega t}\sum_{s_{1}s'_{1}s_{2}s'_{2}}\sum_{\bk_{1}\bp_{1}\bq_{1}}\sum_{\bk_{2}\bp_{2}\bq_{2}} \left(\bv_{\bk{}_{1+}}+\bv_{\bp{}_{1-}}-\bv_{\bk{}_{1-}}-\bv_{\bp{}_{1+}}\right)\left(\bv_{\bk{}_{2+}}+\bv_{\bp{}_{2-}}-\bv_{\bk{}_{2-}}-\bv_{\bp{}_{2+}}\right) \label{eq:INA10}\\
 && \times U(\bk_{1},\bp_{1},\bq_{1}) U(\bk_{2},\bp_{2},\bq_{2}) \langle[c_{\bk{}_{1+},s_{1}}^{\dagger}(t)c_{\bp{}_{1-},s'_{1}}^{\dagger}(t)c_{\bp_{1+}^{\phantom{\dagger}},s'_{1}}(t)c_{\bk_{1-},s_{1}}^{\phantom{\dagger}}(t),c_{\bk{}_{2+},s_{2}}^{\dagger}(0)c_{\bp{}_{2-},s'_{2}}^{\dagger}(0)c_{\bp_{2+},s'_{2}}^{\phantom{\dagger}}(0)c_{\bk_{2-},s{}_{2}}^{\phantom{\dagger}}(0)]\rangle.\nn
\eea
Since \Eq{eq:INA10} already contains a square of the interaction, we can safely 
put $H= H_{0}$ in the evolution operators $e^{iHt}\approx e^{iH_{0}t}$, which leads to 
\bea
c_{\bk{}_{1+},s_{1}}^{\dagger}(t)c_{\bp{}_{1-},s'_{1}}^{\dagger}(t)c_{\bp_{1+},s'_{1}}^{\phantom{\dagger}}(t)c_{\bk_{1-},s_{1}}^{\phantom{\dagger}}(t) = c_{\bk{}_{1+},s_{1}}^{\dagger}c_{\bp{}_{1-},s'_{1}}^{\dagger}c_{\bp_{1+},s'_{1}}^{\phantom{\dagger}}c_{\bk_{1-},s_{1}}^{\phantom{\dagger}} e^{i(\ve_{\bk_{1+}}+\ve_{\bp_{1-}}-\ve_{\bp_{1+}}-\ve_{\bk_{1-}})t}.
\label{eq:t-dep}
\eea
Adding an infinitesimally small imaginary part to $\omega$,
we solve the time integral in \Eq{eq:INA10} as 
\bea
\int_{0}^{\infty} dt  e^{i\left( \omega+ i 0^+ +\ve_{\bk_{1+}}+\ve_{\bp_{1-}}-\ve_{\bp_{1+}}-\ve_{\bk_{1-}}\right)t} = \frac{i}{\omega+\ve_{\bk_{1+}}+\ve_{\bp_{1-}}-\ve_{\bp_{1+}}-\ve_{\bk_{1-}} + i 0^+}.
\eea
Therefore,
\bea
\I \langle\left[{\bf{K}}_1(t),{\bf{K}}_1(0)\right]\rangle_{\omega} && = - \frac{\pi e^{2}}{4}\sum_{s_{1}s'_{1}s_{2}s'_{2}}\sum_{\bk_{1}\bp_{1}\bq_{1}}\sum_{\bk_{2}\bp_{2}\bq_{2}}  \delta(\omega +\ve_{\bk_{1+}}+\ve_{\bp_{1-}}-\ve_{\bp_{1+}}-\ve_{\bk_{1-}}) U(\bk_{1},\bp_{1},\bq_{1}) U(\bk_{2},\bp_{2},\bq_{2}) \nn \\ && \times 
\left(\bv_{\bk{}_{1+}}+\bv_{\bp{}_{1-}}-\bv_{\bk{}_{1-}}-\bv_{\bp{}_{1+}}\right)\left(\bv_{\bk{}_{2+}}+\bv_{\bp{}_{2-}}-\bv_{\bk{}_{2-}}-\bv_{\bp{}_{2+}}\right)
\bigg( \langle \cdots\rangle^\mathrm{I} - \langle \cdots\rangle^\mathrm{II} \bigg),
 \label{eq:INA10-1} 
\eea
where we defined
\bse
\begin{align}
 &\langle \cdots\rangle^\mathrm{I} =\langle c_{\bk{}_{1+},s_{1}}^{\dagger}c_{\bp{}_{1-},s'_{1}}^{\dagger}c_{\bp_{1+}^{\phantom{\dagger}},s'_{1}}c_{\bk_{1-},s_{1}}^{\phantom{\dagger}}c_{\bk{}_{2+},s_{2}}^{\dagger}c_{\bp{}_{2-},s'_{2}}^{\dagger}c_{\bp_{2+},s'_{2}}^{\phantom{\dagger}}c_{\bk_{2-},s{}_{2}}^{\phantom{\dagger}}\rangle,\label{I} \\ 
& \langle \cdots\rangle^\mathrm{II} =\langle c_{\bk{}_{2+},s_{2}}^{\dagger}c_{\bp{}_{2-},s'_{2}}^{\dagger}c_{\bp_{2+},s'_{2}}^{\phantom{\dagger}}c_{\bk_{2-},s{}_{2}}^{\phantom{\dagger}} c_{\bk{}_{1+},s_{1}}^{\dagger}c_{\bp{}_{1-},s'_{1}}^{\dagger}c_{\bp_{1+}^{\phantom{\dagger}},s'_{1}}c_{\bk_{1-},s_{1}}^{\phantom{\dagger}}\rangle.
\label{II} \end{align}
\ese
We now proceed by evaluating the average of the first term in the commutator, $\langle\cdots\rangle^\mathrm{I}$,
using Wick's theorem:
\bea
 \langle\cdots\rangle^\mathrm{I} & =
 -\left\langle c_{\bk{}_{1+},s_{1}}^{\dagger}c_{\bk_{2-},s{}_{2}}^{\phantom{\dagger}}\right\rangle \left\langle c_{\bp_{1-},s'_{1}}^{\dagger}c_{\bp_{2+}s'_{2}}^{\phantom{\dagger}}\right\rangle \left\langle c_{\bp_{1+},s'_{1}}^{\phantom{\dagger}} c_{\bk_{2+},s_{2}}^{\dagger}\right\rangle \left\langle c_{\bk_{1-},s_{1}}^{\phantom{\dagger}} c_{\bp_{2-},s'_{2}}^{\dagger}\right\rangle \nn \\ & 
 +\left\langle c_{\bk_{1+},s_{1}}^{\dagger} c_{\bk_{2-}^{\phantom{\dagger}},s_{2}}\right\rangle \left\langle c_{\bp_{1-},s'_{1}}^{\dagger}c_{\bp_{2+},s'_{2}}^{\phantom{\dagger}} \right\rangle \left\langle c_{\bp_{1+},s'_{1}}^{\phantom{\dagger}} c_{\bp_{2-},s'_{2}}^{\dagger}\right\rangle \left\langle c_{\bk_{1-},s_{1}}^{\phantom{\dagger}} c_{\bk_{2+},s_{2}}^{\dagger}\right\rangle 
 \nn \\& 
 +\left\langle c_{\bk_{1+},s_{1}}^{\dagger}c_{\bp_{2+},s'_{2}}^{\phantom{\dagger}}\right\rangle \left\langle c_{\bp_{1-},s'_{1}}^{\dagger}c_{\bk_{2-},s_{2}}^{\phantom{\dagger}}\right\rangle \left\langle c_{\bp_{1+},s'_{1}}^{\phantom{\dagger}} c_{\bk_{2+},s_{2}}^{\dagger}\right\rangle \left\langle c_{\bk_{1-},s_{1}}^{\phantom{\dagger}} c_{\bp_{2-},s'_{2}}^{\dagger}\right\rangle
 \nn \\ & 
 -\left\langle c_{\bk_{1+},s_{1}}^{\dagger}c_{\bp_{2+},s'_{2}}\right\rangle \left\langle c_{\bp_{1-},s'_{1}}^{\dagger}c_{\bk_{2-},s{}_{2}}^{\phantom{\dagger}}\right\rangle \left\langle c_{\bp_{1+},s'_{1}}^{\phantom{\dagger}}c_{\bp_{2-},s'_{2}}^{\dagger}\right\rangle \left\langle c_{\bk_{1-},s_{1}}^{\phantom{\dagger}} c_{\bk_{2},s_{2}}^{\dagger}\right\rangle.  
 \label{eq:k1k1avg} 
\eea
Using the equal-time averages $\langle c_{{\bf{m}},s_{1}}^{\dagger}c_{{\bf{n}},s{}_{2}}^{\phantom{\dagger}} \rangle = \delta
_{\bf{m},\bf{n}} \delta_{s_1,s_2} n_F(\ve_{\bf{m}})$ and $\langle c_{{\bf{m}},s_{1}}^{\phantom{\dagger}}c_{{\bf{n}},s{}_{2}}^{\dagger} \rangle =  \delta
_{\bf{m},\bf{n}} \delta_{s_1,s_2} \left[1-n_F(\ve_{\bf{m}})\right]$, we sum over over $\bk_2$, $\bp_2$, and all spin indices in \Eq{eq:INA10-1} to obtain
\bea
&&\sum_{s_1,s_2,s'_{1}s'_{2}}\sum_{\bk_{2}\bp_{2}\bq_{2}}  U(\bk_{2},\bp_{2},\bq_{2}) \left(\bv_{\bk{}_{2+}}+\bv_{\bp{}_{2-}}-\bv_{\bk{}_{2-}}-\bv_{\bp{}_{2+}}\right) \langle\cdots\rangle^\mathrm{I} = \left(\bv_{\bk{}_{1+}}+\bv_{\bp{}_{1-}}-\bv_{\bk{}_{1-}}-\bv_{\bp{}_{1+}}\right) \nn \\ &\times& \left[2U\left(
\frac{\bk_1+\bp_1+\bq_1}{2},
\frac{\bk_1+\bp_1-\bq_1}{2} 
\bp_1 - \bk_1\right) 
 - 4U(\bk_1,\bp_1,-\bq_1)-4U(\bp_1,\bk_1,\bq_1)\right. \nn \\ 
  &&\left. + 2U\left(\frac{\bk_1+\bp_1-\bq_1}{2},\frac{\bk_1+\bp_1+\bq_1}{2},\bk_1 - \bp_1\right)\right] 
  n_F(\ve_{\bk_{1+}}) n_F(\ve_{\bp_{1-}}) \left[1-n_F(\ve_{\bk_{1-}})\right] \left[1-n_F(\ve_{\bp_{1+}})\right].
\eea 
All other terms vanish due to incompatible conditions on the electron momenta. Recalling that $U(\bk,\bp,-\bq) = F(\bk) F(\bp)V(-\bq)=F(\bk) F(\bp)V(\bq)=U(\bk,\bp,\bq)$, we simplify the last expression to 
\bea
&&\sum_{s_1,s_2,s'_{1}s'_{2}}\sum_{\bk_{2}\bp_{2}\bq_{2}}  U(\bk_{2},\bp_{2},\bq_{2}) \left(\bv_{\bk{}_{2+}}+\bv_{\bp{}_{2-}}-\bv_{\bk{}_{2-}}-\bv_{\bp{}_{2+}}\right) \langle\cdots\rangle^\mathrm{I} = -8\left(\bv_{\bk{}_{1+}}+\bv_{\bp{}_{1-}}-\bv_{\bk{}_{1-}}-\bv_{\bp{}_{1+}}\right) \nn \\ &&\times \bigg[ U(\bk_1,\bp_1,\bq_1)-\frac{1}{2} U\bigg(\frac{\bk_1+\bp_1+\bq_1}{2},\frac{\bk_1+\bp_1-\bq_1}{2},\bp_1 - \bk_1\bigg) \bigg]
n_F(\ve_{\bk_{1+}}) n_F(\ve_{\bp_{1-}}) \left[1-n_F(\ve_{\bk_{1-}})\right] \left[1-n_F(\ve_{\bp_{1+}})\right].\label{k1k1comm1}
\eea
Likewise, we obtain for the second, $\langle\cdots\rangle^\mathrm{II}$ term  in \Eq{eq:INA10-1}:
\bea
&&\sum_{s_1,s_2,s'_{1}s'_{2}}\sum_{\bk_{2}\bp_{2}\bq_{2}}  U(\bk_{2},\bp_{2},\bq_{2}) \left(\bv_{\bk{}_{2+}}+\bv_{\bp{}_{2-}}-\bv_{\bk{}_{2-}}-\bv_{\bp{}_{2+}}\right) \langle\cdots\rangle^\mathrm{II}  =- 8\left(\bv_{\bk{}_{1+}}+\bv_{\bp{}_{1-}}-\bv_{\bk{}_{1-}}-\bv_{\bp{}_{1+}}\right) \nn \\ &&\times \left[ U(\bk_1,\bp_1,\bq_1)-\frac{1}{2} U\bigg(\frac{\bk_1+\bp_1+\bq_1}{2},\frac{\bk_1+\bp_1-\bq_1}{2},\bp_1 - \bk_1\bigg) \right] n_F(\ve(_{\bk_{1-}}) n_F(\ve_{\bp_{1+}}) \left[1-n_F(\ve_{\bk_{1+}})\right] \left[1-n_F(\ve_{\bp_{1-}})\right].\label{k1k1comm2}
\eea
Substituting Eqs.~\eqref{k1k1comm1} and \eqref{k1k1comm2} into \Eq{eq:INA10-1}, using the property $1-n_F(\ve) = {e^{\ve/T} }n_F(\ve)$, and relabeling $\bk_1\to\bk$, $\bp_1\to \bp$, $\bq_1\to\bq$, we get
\bea
\I \langle\left[{\bf{K}}_1(t),{\bf{K}}_1(0)\right]\rangle_{\omega} & =&{2\pi e^{2}} (1-e^{-\omega/T})
\sum_{\bk\bp\bq} \delta(\omega +\ve_{\bk_{+}}+\ve_{\bp_{-}}-\ve_{\bp_{+}}-\ve_{\bk_{-}}) 
\left(\bv_{\bk{}_{+}}+\bv_{\bp{}_{-}}-\bv_{\bk{}_{-}}-\bv_{\bp{}_{+}}\right)^2\nn \\ & \times &
U(\bk,\bp,\bq) \bigg[ U(\bk,\bp,\bq)-\frac{1}{2} U\bigg(\frac{\bk+\bp+\bq}{2},\frac{\bk+\bp-\bq}{2},\bp - \bk\bigg) \bigg] M(\bk,\bp,\bq),\label{K1K1final} 
\eea
where $M(\bk,\bp,\bq)$ is given by Eq.~\eqref{M}. The second term in $[\dots]$  in the last expression is the exchange part of the interaction, which is small compared to the first (direct) term for our case of a long-range $U(\bq)$. Neglecting the exchange term and substituting \Eq{K1K1final} into \Eq{k1k1}, we obtain \Eq{sigma1} of the main text.

We now turn to  \Eq{sigma2}. Following the same steps as above 
up to \Eq{eq:k1k1avg}, we arrive at
\bea
&&\I \langle\left[{\bf{K}}_2(t),{\bf{K}}_2(0)\right]\rangle_{\omega}  = -{\pi e^{2}}\sum_{s_{1}s'_{1}s_{2}s'_{2}}\sum_{\bk_{1}\bp_{1}\bq_{1}}\sum_{\bk_{2}\bp_{2}\bq_{2}}  \delta(\omega +\ve_{\bk_{1+}}+\ve_{\bp_{1-}}-\ve_{\bp_{1+}}-\ve_{\bk_{1-}})  \left(\boldsymbol{\nabla}_{\bk_1}+\bn_{\bp_1} \right) U(\bk_{1},\bp_{1},\bq_{1})   \nn \\ & & \times \left(\boldsymbol{\nabla}_{\bk_2}+\bn_{\bp_2} \right)U(\bk_{2},\bp_{2},\bq_{2})
\left(\ve_{\bk{}_{1+}}+\ve_{\bp{}_{1-}}-\ve_{\bk{}_{1-}}-\ve_{\bp{}_{1+}}\right)\left(\ve_{\bk{}_{2+}}+\ve_{\bp{}_{2-}}-\ve_{\bk{}_{2-}}-\ve_{\bp{}_{2+}}\right)
\bigg( \langle \cdots\rangle^\mathrm{I} - \langle \cdots\rangle^\mathrm{II} \bigg),
 \label{eq:INA10_k2k2} 
\eea
where $\langle\dots\rangle^\mathrm{I}$ and $\langle\dots\rangle^{\mathrm{II}}$ are the same as in Eqs.~\eqref{I} and \eqref{II}. Summing the 
$\langle\dots\rangle^\mathrm{I}$ term in \Eq{eq:INA10_k2k2} over $\bk_2$, $\bp_2$, and over spins, we find
\bea
&&\sum_{s_1,s_2,s'_{1}s'_{2}}\sum_{\bk_{2}\bp_{2}\bq_{2}} \left(\boldsymbol{\nabla}_{\bk_2}+\bn_{\bp_2} \right)  U(\bk_{2},\bp_{2},\bq_{2}) \left(\ve_{\bk{}_{2+}}+\ve_{\bp{}_{2-}}-\ve_{\bk{}_{2-}}-\ve_{\bp{}_{2+}}\right) \langle\cdots\rangle^\mathrm{I}  =  - 8\left(\ve_{\bk{}_{1+}}+\ve_{\bp{}_{1-}}-\ve_{\bk{}_{1-}}-\ve_{\bp{}_{1+}}\right) \nn\\
&&\times\left[ \left(\boldsymbol{\nabla}_{\bk_1}+\bn_{\bp_1} \right) U(\bk_{1},\bp_{1},\bq_{1}) - \frac{1}{2}\left(\boldsymbol{\nabla}_{\bk_2}+\bn_{\bp_2} \right) U(\bk_{2},\bp_{2},
\bp_1-\bk_1) \bigg|_{\bk_2 = \frac{\bk_1+\bp_1+\bq_1}{2}, \bp_2 = \frac{\bk_1+\bp_1-\bq_1}{2}}
\right]   \nn \\ 
 &&\times 
  n_F(\ve_{\bk_{1+}}) n_F(\ve_{\bp_{1-}}) \left[1-n_F(\ve_{\bk_{1-}})\right] \left[1-n_F(\ve_{\bp_{1+}})\right]. \label{eq:k2k2-avg1}
\eea 
Following similar steps for the $\langle\dots\rangle^\mathrm{II}$ term in \Eq{eq:INA10_k2k2} and using the same property of the Fermi functions, 
we arrive at 
\bea
&&\I \langle\left[{\bf{K}}_2(t),{\bf{K}}_2(0)\right]\rangle_{\omega}  ={8\pi e^{2}} (1-e^{- \omega/T})
\sum_{\bk_{1}\bp_{1}\bq_{1}} \delta(\omega +\ve_{\bk_{1+}}+\ve_{\bp_{1-}}-\ve_{\bp_{1+}}-\ve_{\bk_{1-}}) 
\left(\ve_{\bk{}_{1+}}+\ve_{\bp{}_{1-}}-\ve_{\bk{}_{1-}}-\ve_{\bp{}_{1+}}\right)^2\nn \\ && \times \left(\boldsymbol{\nabla}_{\bk_1}+\bn_{\bp_1} \right)
U(\bk_{1},\bp_{1},\bq_{1}) \bigg[\left(\boldsymbol{\nabla}_{\bk_1}+\bn_{\bp_1} \right) U(\bk_1,\bp_1,\bq_1) - \frac{1}{2}\left(\boldsymbol{\nabla}_{\bk_2}+\bn_{\bp_2} \right) U(\bk_{2},\bp_{2},\bq_{2}) \bigg|_{\bk_2 = \frac{\bk_1+\bp_1+\bq_1}{2}, \bp_2 = \frac{\bk_1+\bp_1-\bq_1}{2}, \bq_2 = \bp_1 -\bk_1} 
\bigg]
\nn \\
&&\times   n_F(\ve_{\bk_{1+}}) n_F(\ve_{\bp_{1-}}) \left[1-n_F(\ve_{\bk_{1-}})\right] \left[1-n_F(\ve_{\bp_{1+}})\right].
 \label{eq:INA10-k2k2-2} 
\eea
Neglecting the second (exchange) term in $[\dots]$ in the last expression and substituting the result into \Eq{k2k2}, we obtain \Eq{sigma2} of the main text.

Finally, we turn to Eq.~\eqref{k1k2} which contains the cross-correlators, $\I \langle\left[{\bf{K}}_1(t),{\bf{K}}_2(0)\right]\rangle_{\pm\omega}$. 
The same steps as before lead us to
\bea
&&\I \langle\left[{\bf{K}}_1(t),{\bf{K}}_2(0)\right]\rangle_{\omega} =  \frac{\pi e^{2}}{2}\sum_{s_{1}s'_{1}s_{2}s'_{2}}\sum_{\bk_{1}\bp_{1}\bq_{1}}\sum_{\bk_{2}\bp_{2}\bq_{2}}  \delta(\omega +\ve_{\bk_{1+}}+\ve_{\bp_{1-}}-\ve_{\bp_{1+}}-\ve_{\bk_{1-}}) U(\bk_{1},\bp_{1},\bq_{1})  \left(\bv_{\bk{}_{1+}}+\bv_{\bp{}_{1-}}-\bv_{\bk{}_{1-}}-\bv_{\bp{}_{1+}}\right) \nn \\ &&  \cdot 
 \left(\boldsymbol{\nabla}_{\bk_2}+\bn_{\bp_2} \right) U(\bk_{2},\bp_{2},\bq_{2}) \left(\ve_{\bk{}_{2+}}+\ve_{\bp{}_{2-}}-\ve_{\bk{}_{2-}}-\ve_{\bp{}_{2+}}\right)
\bigg( \langle \cdots\rangle^\mathrm{I} - \langle \cdots\rangle^\mathrm{II} \bigg) 
\label{eq:INA10-k1k2} 
\eea 
and 
\bea
&&\I \langle\left[{\bf{K}}_1(t),{\bf{K}}_2(0)\right]\rangle_{\omega}  = {4\pi e^{2}}\omega(1-e^{- \omega/T})
\sum_{\bk_{1}\bp_{1}\bq_{1}}  \delta(\omega +\ve_{\bk_{1+}}+\ve_{\bp_{1-}}-\ve_{\bp_{1+}}-\ve_{\bk_{1-}}) \left(\bv_{\bk{}_{1+}}+\bv_{\bp{}_{1-}}-\bv_{\bk{}_{1-}}-\bv_{\bp{}_{1+}}\right)  \nn \\ 
&& \cdot \bigg[ \left(\boldsymbol{\nabla}_{\bk_1}+\bn_{\bp_1} \right) U(\bk_{1},\bp_{1},\bq_{1})
- \frac{1}{2}\left(\boldsymbol{\nabla}_{\bk_2}+\bn_{\bp_2} \right) U(\bk_{2},\bp_{2}, \bp_1 -\bk_1) \bigg|_{\bk_2 = \frac{\bk_1+\bp_1+\bq_1}{2}, \bp_2 = \frac{\bk_1+\bp_1-\bq_1}{2}}  \bigg]U(\bk_{1},\bp_{1},\bq_{1})\nn\\
&&\times  n_F(\ve_{\bk_{1+}}) n_F(\ve_{\bp_{1-}}) \left[1-n_F(\ve_{\bk_{1-}})\right] \left[1-n_F(\ve_{\bp_{1+}})\right],
\label{eq:INA10-k1k2} 
\eea 
where we used the energy-conserving delta-function to express the energy difference in terms of $\omega$. Replacing $\omega\longrightarrow -\omega$, relabeling $\bk_{1}\leftrightarrow\bp_1$, and taking into account that $U(\bk,\bp,\bq)=U(\bp,\bk,\bq)$, we find 
\bea
&&\I \langle\left[{\bf{K}}_1(t),{\bf{K}}_2(0)\right]\rangle_{\omega}  = {4\pi e^{2}}
\omega(1-e^{ \omega/T})\sum_{\bk_{1}\bp_{1}\bq_{1}}  \delta(\omega +\ve_{\bk_{1+}}+\ve_{\bp_{1-}}-\ve_{\bp_{1+}}-\ve_{\bk_{1-}})
\left(\bv_{\bk{}_{1+}}+\bv_{\bp{}_{1-}}-\bv_{\bk{}_{1-}}-\bv_{\bp{}_{1+}}\right)  \nn \\ 
&& \cdot \bigg[ \left(\boldsymbol{\nabla}_{\bk_1}+\bn_{\bp_1} \right) U(\bk_{1},\bp_{1},\bq_{1})
- \frac{1}{2}\left(\boldsymbol{\nabla}_{\bk_2}+\bn_{\bp_2} \right) U(\bk_{2},\bp_{2}, \bp_1 -\bk_1) \bigg|_{\bk_2 = \frac{\bk_1+\bp_1+\bq_1}{2}, \bp_2 = \frac{\bk_1+\bp_1-\bq_1}{2}}  \bigg]U(\bk_{1},\bp_{1},\bq_{1})\nn\\
&&\times  n_F(\ve_{\bp_{1+}}) n_F(\ve_{\bk_{1-}}) \left[1-n_F(\ve_{\bp_{1-}})\right] \left[1-n_F(\ve_{\bk_{1+}})\right].
\label{eq:INA10-k1k2_2} 
\eea 
Now we apply the identity 
\bea
\left(1-e^{\omega/T}\right)n_F(\ve_{\bk_{1+}}) n_F(\ve_{\bp_{1-}}) \left[1-n_F(\ve_{\bk_{1-}})\right] \left[1-n_F(\ve_{\bp_{1+}})\right]=
-\left(1-e^{-\omega/T}\right)
 n_F(\ve_{\bk_{1-}}) n_F(\ve_{\bp_{1+}}) \left[1-n_F(\ve_{\bk_{1+}})\right] \left[1-n_F(\ve_{\bp_{1-}})\right]
  \label{ident}
\eea
to get 
\bea
\I \langle\left[{\bf{K}}_1(t),{\bf{K}}_2(0)\right]\rangle_{\omega} =
-\I \langle\left[{\bf{K}}_1(t),{\bf{K}}_2(0)\right]\rangle_{-\omega}.
\eea
According to \Eq{k1k2} we then have 
\bea
\sigma'_{12}(\omega,T)=\frac{g^2}{\omega^3}\im\,
\la
\left[{\bf K}_1(t)\stackrel{\cdot}{,}{\bf K}_2(0)\right]
\ra_\omega.\label{twice}\eea
Substituting \Eq{eq:INA10-k1k2} into \Eq{twice} and neglecting the exchange term, we obtain \Eq{Sigma} of the main text.


\section{Properties of $\sigma_{12}'(\omega,T)$}\label{sec:app-C}

In Sec. \ref{sec:concaveA}, we encountered the following expression for the cross-term in the optical conductivity [Eq. \eqref{sigma12_1M}]:
\bea
\label{sigma12_1}
&&\sigma'_{12}(\omega,T)=
 e^2g^2\frac{\pi}{(2\pi)^4\omega^2} (1-e^{-\omega/T}) \int \frac{d^2q}{(2\pi)^2} V^2(\bq) \int^{+\infty}_{-\infty} d \ve_\bk \int^{+\infty}_{-\infty} d\ve_\bp\int^{+\infty}_{-\infty} d\Omega
 \oint \frac{d\ell_\bk}{\varv_\bk}
 \oint \frac{d\ell_\bp}{\varv_\bp}\nn\\
&&\times  ({\bf w}\cdot\Delta \bv)   
n_\mathrm{F}(\ve_{\bk}) n_\mathrm{F}(\ve_\bp ) \left[1-n_\mathrm{F}(\ve_{\bk}+\Omega)\right]
\left[1-n_\mathrm{F}(\ve_\bp-\Omega+\omega)\right]
\delta(\Omega-\ve_{\bk-\bq}+\ve_{\bk} )  
\delta( \Omega-\omega+\ve_{\bp+\bq}-\ve_{\bp}). 
\eea
To check that $\sigma_{12}'(\omega,T)$ is even in $\omega$, we flip the sign of frequency ($\omega\to-\omega$), and relabel $\bk-\bq\leftrightarrow\bp$ and $\bp+\bq\leftrightarrow\bk$. (Note that the last transformation results in $\Delta\bv\to-\Delta\bv$.) Therefore,
\bea
\label{sigma12_3}
&&\sigma'_{12}(-\omega,T) = - 
 e^2g^2\frac{\pi}{(2\pi)^4\omega^2} (1-e^{\omega/T}) \int \frac{d^2q}{(2\pi)^2} V^2(\bq) \int^{+\infty}_{-\infty} d \ve_\bk \int^{+\infty}_{-\infty} d\ve_\bp\int^{+\infty}_{-\infty} d\Omega
 \oint \frac{d\ell_\bk}{\varv_\bk}
 \oint \frac{d\ell_\bp}{\varv_\bp}\nn\\
&&\times  ({\bf w}\cdot\Delta\bv)   
n_\mathrm{F}(\ve_{\bp+\bq})n_\mathrm{F}(\ve_{\bk-\bq}) 
\left[1-n_\mathrm{F}(\ve_{\bp+\bq}+\Omega)\right]\left[1-n_\mathrm{F}(\ve_{\bk-\bq}-\Omega-\omega)\right]
\delta(\Omega-\ve_{\bp}+\ve_{\bp+\bq} )
\delta(\Omega+\omega+\ve_{\bk}-\ve_{\bk-\bq}).
\eea 
Next, we shift the integration variable as $\Omega+\omega\to \Omega$ and eliminate $\ve_{\bk-\bq}$ and $\ve_{\bp+\bq}$ in favor of $\ve_\bk$ and $\ve_\bp$, using the delta-functions. This gives:
 \bea
\label{sigma12_4}
&&\sigma'_{12}(-\omega,T) 
=-e^2g^2\frac{\pi}{(2\pi)^4\omega^2} (1-e^{\omega/T}) \int \frac{d^2q}{(2\pi)^2} V^2(\bq) \int^{+\infty}_{-\infty} d \ve_\bk \int^{+\infty}_{-\infty} d\ve_\bp\int^{+\infty}_{-\infty} d\Omega
 \oint \frac{d\ell_\bk}{\varv_\bk}
 \oint \frac{d\ell_\bp}{\varv_\bp}\nn\\
&&\times({\bf w}\cdot\Delta\bv)   
n_\mathrm{F}(\ve_{\bp}-\Omega+\omega) n_\mathrm{F}(\ve_{\bk}+\Omega) 
\left[1-n_\mathrm{F}(\ve_{\bp})\right]\left[1-n_\mathrm{F}(\ve_{\bk})\right]
\delta(\Omega-\omega-\ve_{\bp}+\ve_{\bp+\bq} )  \delta( \Omega
+\ve_{\bk}-\ve_{\bk-\bq}).
\eea
Now the delta functions are the same as in Eq. \eqref{sigma12_1}. Further, we rewrite the product of the Fermi functions in the last equation as
\bea
&&n_\mathrm{F}(\ve_{\bp}-\Omega+\omega
) n_\mathrm{F}(\ve_{\bk}+\Omega) 
\left[1-n_\mathrm{F}(\ve_{\bp})\right]\left[1-n_\mathrm{F}(\ve_{\bk})\right]\nn\\
&&=n_\mathrm{F}(\ve_{\bp}-\Omega+\omega
) n_\mathrm{F}(\ve_{\bk}+\Omega) n_F(\ve_\bp)n_F(\ve_\bk) e^{\beta(\ve_\bp+\ve_\bk)}\nn\\
&&=n_\mathrm{F}(\ve_{\bp}-\Omega+\omega
)n_\mathrm{F}(\ve_{\bk}+\Omega)n_F(\ve_\bp)n_F(\ve_\bk) e^{\beta(\ve_\bp+\ve_\bk)} e^{\beta(\ve_\bp-\Omega+\omega)}e^{-\beta(\ve_\bp-\Omega+\omega)}e^{\beta(\ve_\bk+\Omega)}e^{-\beta(\ve_\bk+\Omega)}\nn\\
&&=\left[1-n_\mathrm{F}(\ve_{\bp}-\Omega+\omega
)\right]\left[1-n_\mathrm{F}(\ve_{\bk}+\Omega) \right]n_F(\ve_\bp)n_F(\ve_\bk) e^{-\beta\omega}.
\eea
Substituting this back into Eq.~\eqref{sigma12_1}, we see that indeed $\sigma_{12}'(-\omega,T)=\sigma_{12}'(\omega,T)$, as it should.

Nevertheless, if we neglect the frequencies in the delta functions and use Eq.~\eqref{energy_int} for the energy integrals, we get an \emph{odd} function of frequency, i.e.,  $\sigma'_{12}(\omega,T)\propto \omega(1+4\pi^2 T^2/\omega^2)$. Therefore, one cannot neglect the frequencies in this case. Note that $\sigma'_{12}(\omega,T)$ is even only because $\Delta\bv$ changes sign on the transformations employed. Now, let's expand the dispersions in the delta-functions in Eq.~\eqref{sigma12_1} to $O(q)$ but keep the frequencies in there, and see if the evenness of $\sigma'_{12}(\omega,T)$ is preserved by these simplifications. Performing the steps indicated above, we obtain
\bea
\label{sigma12_6}
&&\sigma'_{12}(\omega,T)=  
 e^2g^2\frac{\pi}{(2\pi)^4\omega^2} (1-e^{-\omega/T}) \int \frac{d^2q}{(2\pi)^2} V^2(\bq) \int^{+\infty}_{-\infty} d \ve_\bk \int^{+\infty}_{-\infty} d\ve_\bp\int^{+\infty}_{-\infty} d\Omega
 \oint \frac{d\ell_\bk}{\varv_\bk}
 \oint \frac{d\ell_\bp}{\varv_\bp}\nn\\
&&\times({\bf w}\cdot\Delta \bv)
n_\mathrm{F}(\ve_{\bk}) n_\mathrm{F}(\ve_\bp) 
\left[1-n_\mathrm{F}(\ve_{\bk}+\Omega)\right]
\left[1-n_\mathrm{F}(\ve_\bp-\Omega+\omega)\right]
\delta(\Omega+\bv^F_\bk\cdot\bq)  \delta( \Omega-\omega+\bv^F_\bp\cdot\bq),
\eea
where the subscript $F$ indicates that the velocities are taken right on the FS, i.e., they do not depend on $\ve_\bk$ and $\ve_\bp$. On $\omega\to-\omega$, the last expression becomes
\bea
\label{sigma12_7}
&&\sigma'_{12}(-\omega,T)=
 e^2g^2\frac{\pi}{(2\pi)^4\omega^2} (1-e^{\omega/T}) \int \frac{d^2q}{(2\pi)^2} V^2(\bq) \int^{+\infty}_{-\infty} d \ve_\bk \int^{+\infty}_{-\infty} d\ve_\bp\int^{+\infty}_{-\infty} d\Omega
 \oint \frac{d\ell_\bk}{\varv_\bk}
 \oint \frac{d\ell_\bp}{\varv_\bp}\nn\\
&&\times  ({\bf w}\cdot\Delta \bv^F)
n_\mathrm{F}(\ve_{\bk}) n_\mathrm{F}(\ve_\bp) \left[1-n_\mathrm{F}(\ve_{\bk}+\Omega)\right]\left[1-n_\mathrm{F}(\ve_\bp-\Omega-\omega)\right]
\delta(\Omega+\bv^F_\bk\cdot\bq)  
\delta( \Omega+\omega+\bv^F_\bp\cdot\bq).
\eea
Now we perform the following sequence of transformations: relabel $\bk\leftrightarrow\bp$ (note that $\Delta\bv$ does not change its sign on this transformation), 
shift the variable as $\Omega+\omega\to\Omega$, and replace 
$\ve_{\bk,\bp}\to-\ve_{\bk,\bp}$ (note that the velocities are not affected by this transformation). Then we obtain 
\bea
\label{sigma12_10}
&&\sigma'_{12}(-\omega,T)=
 e^2g^2\frac{\pi}{(2\pi)^4\omega^2} (1-e^{\omega/T}) \int \frac{d^2q}{(2\pi)^2} V^2(\bq) \int^{+\infty}_{-\infty} d \ve_\bk \int^{+\infty}_{-\infty} d\ve_\bp\int^{+\infty}_{-\infty} d\Omega
 \oint \frac{d\ell_\bk}{\varv_\bk}
 \oint \frac{d\ell_\bp}{\varv_\bp}\nn\\
&&\times({\bf w}\cdot\Delta \bv^F)
n_\mathrm{F}(-\ve_{\bp}) n_\mathrm{F}(-\ve_\bk ) 
\left[1-n_\mathrm{F}(-\ve_{\bp}+\Omega-\omega)\right]
\left[1-n_\mathrm{F}(-\ve_\bk-\Omega)\right]
\delta(\Omega-\omega+\bv^F_\bp\cdot\bq)  
\delta( \Omega+\bv^F_\bk\cdot\bq).
\eea
The rest of the transformations affect only the Fermi functions:
\bea
&&n_\mathrm{F}(-\ve_{\bp}) n_\mathrm{F}(-\ve_\bk ) 
\left[1-n_\mathrm{F}(-\ve_{\bp}+\Omega-\omega)\right]
\left[1-n_\mathrm{F}(-\ve_\bk-\Omega)\right]\nn\\
&&=[1-n_F(\ve_\bp)][1-n_F(\ve_\bk)]n_F(\ve_\bp-\Omega+\omega) n_F(\ve_\bk+\Omega)\nn\\
&&=n_F(\ve_\bp)n_F(\ve_\bk)e^{\beta(\ve_\bk+\ve_\bp)}n_F(\ve_\bp-\Omega+\omega)e^{\beta(\ve_\bp-\Omega+\omega)}e^{-\beta(\ve_\bp-\Omega+\omega)}n_F(\ve_\bk+\Omega)e^{\beta(\ve_\bk+\Omega)}e^{-\beta(\ve_\bk+\Omega)}\nn\\
&&=n_F(\ve_\bp)n_F(\ve_\bk)[1-n_F(\ve_\bp-\Omega+\omega)][1-n_F(\ve_\bk+\Omega]e^{-\beta\omega}.\label{nF}
\eea
Substituting \Eq{nF} into \Eq{sigma12_10} yields $\sigma'_{12}(-\omega,T)=-\sigma'_{12}(\omega,T)$, which is incorrect.

Let's summarize: projecting the momenta onto the FS yields $\sigma'_{12}(\omega,T)=A\omega$ (at $T=0$). However, such a term is not allowed by time-reversal
symmetry; thus we must have $A=0$, regardless of the shape of the FS. Therefore, even for a concave FS, one needs to expand $\Delta\bv$ in deviations from the FS, as done for the isotropic case. That should give $\sigma'_{12}(\omega,T)=B\omega^2$, which is subleading to the $\sigma'_1(\omega,T)$.
\end{widetext}


\bibliography{optcondIN}

\begin{thebibliography}{30}%
\makeatletter
\providecommand \@ifxundefined [1]{%
 \@ifx{#1\undefined}
}%
\providecommand \@ifnum [1]{%
 \ifnum #1\expandafter \@firstoftwo
 \else \expandafter \@secondoftwo
 \fi
}%
\providecommand \@ifx [1]{%
 \ifx #1\expandafter \@firstoftwo
 \else \expandafter \@secondoftwo
 \fi
}%
\providecommand \natexlab [1]{#1}%
\providecommand \enquote  [1]{``#1''}%
\providecommand \bibnamefont  [1]{#1}%
\providecommand \bibfnamefont [1]{#1}%
\providecommand \citenamefont [1]{#1}%
\providecommand \href@noop [0]{\@secondoftwo}%
\providecommand \href [0]{\begingroup \@sanitize@url \@href}%
\providecommand \@href[1]{\@@startlink{#1}\@@href}%
\providecommand \@@href[1]{\endgroup#1\@@endlink}%
\providecommand \@sanitize@url [0]{\catcode `\\12\catcode `\$12\catcode
  `\&12\catcode `\#12\catcode `\^12\catcode `\_12\catcode `\%12\relax}%
\providecommand \@@startlink[1]{}%
\providecommand \@@endlink[0]{}%
\providecommand \url  [0]{\begingroup\@sanitize@url \@url }%
\providecommand \@url [1]{\endgroup\@href {#1}{\urlprefix }}%
\providecommand \urlprefix  [0]{URL }%
\providecommand \Eprint [0]{\href }%
\providecommand \doibase [0]{http://dx.doi.org/}%
\providecommand \selectlanguage [0]{\@gobble}%
\providecommand \bibinfo  [0]{\@secondoftwo}%
\providecommand \bibfield  [0]{\@secondoftwo}%
\providecommand \translation [1]{[#1]}%
\providecommand \BibitemOpen [0]{}%
\providecommand \bibitemStop [0]{}%
\providecommand \bibitemNoStop [0]{.\EOS\space}%
\providecommand \EOS [0]{\spacefactor3000\relax}%
\providecommand \BibitemShut  [1]{\csname bibitem#1\endcsname}%
\let\auto@bib@innerbib\@empty
\bibitem [{\citenamefont {Basov}\ and\ \citenamefont
  {Timusk}(2005)}]{basov:2005}%
  \BibitemOpen
  \bibfield  {author} {\bibinfo {author} {\bibfnamefont {D.~N.}\ \bibnamefont
  {Basov}}\ and\ \bibinfo {author} {\bibfnamefont {T.}~\bibnamefont {Timusk}},\
  }\bibfield  {title} {\enquote {\bibinfo {title} {Electrodynamics of
  high-${T}_{c}$ superconductors},}\ }\href {\doibase
  10.1103/RevModPhys.77.721} {\bibfield  {journal} {\bibinfo  {journal} {Rev.
  Mod. Phys.}\ }\textbf {\bibinfo {volume} {77}},\ \bibinfo {pages} {721--779}
  (\bibinfo {year} {2005})}\BibitemShut {NoStop}%
\bibitem [{\citenamefont {{Basov}}\ \emph {et~al.}(2011)\citenamefont
  {{Basov}}, \citenamefont {{Averitt}}, \citenamefont {{van der Marel}},
  \citenamefont {{Dressel}},\ and\ \citenamefont {{Haule}}}]{basov:2011}%
  \BibitemOpen
  \bibfield  {author} {\bibinfo {author} {\bibfnamefont {D.~N.}\ \bibnamefont
  {{Basov}}}, \bibinfo {author} {\bibfnamefont {R.~D.}\ \bibnamefont
  {{Averitt}}}, \bibinfo {author} {\bibfnamefont {D.}~\bibnamefont {{van der
  Marel}}}, \bibinfo {author} {\bibfnamefont {M.}~\bibnamefont {{Dressel}}}, \
  and\ \bibinfo {author} {\bibfnamefont {K.}~\bibnamefont {{Haule}}},\
  }\bibfield  {title} {\enquote {\bibinfo {title} {{Electrodynamics of
  correlated electron materials}},}\ }\href {\doibase
  10.1103/RevModPhys.83.471} {\bibfield  {journal} {\bibinfo  {journal} {Rev.
  Mod. Phys.}\ }\textbf {\bibinfo {volume} {83}},\ \bibinfo {pages} {471--542}
  (\bibinfo {year} {2011})}\BibitemShut {NoStop}%
\bibitem [{\citenamefont {Maslov}\ and\ \citenamefont
  {Chubukov}(2017)}]{maslov:2017b}%
  \BibitemOpen
  \bibfield  {author} {\bibinfo {author} {\bibfnamefont {Dmitrii~L.}\
  \bibnamefont {Maslov}}\ and\ \bibinfo {author} {\bibfnamefont {Andrey~V.}\
  \bibnamefont {Chubukov}},\ }\bibfield  {title} {\enquote {\bibinfo {title}
  {Optical response of correlated electron systems},}\ }\href
  {http://iopscience.iop.org/article/10.1088/1361-6633/80/2/026503} {\bibfield
  {journal} {\bibinfo  {journal} {Rep. Prog. Phys.}\ }\textbf {\bibinfo
  {volume} {80}},\ \bibinfo {pages} {026503} (\bibinfo {year}
  {2017})}\BibitemShut {NoStop}%
\bibitem [{\citenamefont {Armitage}(2018)}]{Armitage:2018}%
  \BibitemOpen
  \bibfield  {author} {\bibinfo {author} {\bibfnamefont {N.~P.}\ \bibnamefont
  {Armitage}},\ }\href@noop {} {\enquote {\bibinfo {title} {Electrodynamics of
  correlated electron systems},}\ } (\bibinfo {year} {2018}),\ \Eprint
  {http://arxiv.org/abs/0908.1126} {arXiv:0908.1126} \BibitemShut {NoStop}%
\bibitem [{\citenamefont {Tanner}(2019)}]{Tanner:book}%
  \BibitemOpen
  \bibfield  {author} {\bibinfo {author} {\bibfnamefont {D.~B.}\ \bibnamefont
  {Tanner}},\ }\href@noop {} {\emph {\bibinfo {title} {{Optical Effects in
  Solids}}}}\ (\bibinfo  {publisher} {Cambridge University Press, Cambridge},\
  \bibinfo {year} {2019})\BibitemShut {NoStop}%
\bibitem [{\citenamefont {Landau}\ and\ \citenamefont
  {Pomeranchuk}(1936)}]{Landau:1936}%
  \BibitemOpen
  \bibfield  {author} {\bibinfo {author} {\bibfnamefont {L.D.}\ \bibnamefont
  {Landau}}\ and\ \bibinfo {author} {\bibfnamefont {I.~Ya.}\ \bibnamefont
  {Pomeranchuk}},\ }\bibfield  {title} {\enquote {\bibinfo {title} {On the
  properties of metals at very low temperatures},}\ }\href@noop {} {\bibfield
  {journal} {\bibinfo  {journal} {Ph. Zs. Sowjet.}\ }\textbf {\bibinfo {volume}
  {10}},\ \bibinfo {pages} {649} (\bibinfo {year} {1936})}\BibitemShut
  {NoStop}%
\bibitem [{\citenamefont {Ter-Haar}(1965)}]{Landau:collected}%
  \BibitemOpen
  \bibinfo {editor} {\bibfnamefont {D.}~\bibnamefont {Ter-Haar}},\ ed.,\
  \href@noop {} {\emph {\bibinfo {title} {{Collected Papers of L. D.
  Landau}}}}\ (\bibinfo  {publisher} {Oxford,Pergamon},\ \bibinfo {year}
  {1965})\BibitemShut {NoStop}%
\bibitem [{\citenamefont {Lifshitz}\ and\ \citenamefont
  {Pitaevskii}(1981)}]{physkin}%
  \BibitemOpen
  \bibfield  {author} {\bibinfo {author} {\bibfnamefont {E.~M.}\ \bibnamefont
  {Lifshitz}}\ and\ \bibinfo {author} {\bibfnamefont {L.~P.}\ \bibnamefont
  {Pitaevskii}},\ }\href@noop {} {\emph {\bibinfo {title} {Physical Kinetics,
  Course of Theoretical Physics, v. X}}}\ (\bibinfo  {publisher}
  {Butterworth-Heinemann, Burlington},\ \bibinfo {year} {1981})\BibitemShut
  {NoStop}%
\bibitem [{\citenamefont {{Baber}}(1937)}]{baber:1937}%
  \BibitemOpen
  \bibfield  {author} {\bibinfo {author} {\bibfnamefont {W.~G.}\ \bibnamefont
  {{Baber}}},\ }\bibfield  {title} {\enquote {\bibinfo {title} {{The
  Contribution to the Electrical Resistance of Metals from Collisions between
  Electrons}},}\ }\href {\doibase 10.1098/rspa.1937.0027} {\bibfield  {journal}
  {\bibinfo  {journal} {Proc. Royal Soc. London A}\ }\textbf {\bibinfo {volume}
  {158}},\ \bibinfo {pages} {383--396} (\bibinfo {year} {1937})}\BibitemShut
  {NoStop}%
\bibitem [{\citenamefont {Gurzhi}(1959)}]{gurzhi:1959}%
  \BibitemOpen
  \bibfield  {author} {\bibinfo {author} {\bibfnamefont {R.~N.}\ \bibnamefont
  {Gurzhi}},\ }\bibfield  {title} {\enquote {\bibinfo {title} {{Mutual Electron
  Correlations in Metal Optics}},}\ }\href@noop {} {\bibfield  {journal}
  {\bibinfo  {journal} {Sov. Phys.--JETP}\ }\textbf {\bibinfo {volume} {35}},\
  \bibinfo {pages} {673} (\bibinfo {year} {1959})}\BibitemShut {NoStop}%
\bibitem [{\citenamefont {Pal}\ \emph {et~al.}(2012{\natexlab{a}})\citenamefont
  {Pal}, \citenamefont {Yudson},\ and\ \citenamefont {Maslov}}]{pal:2012b}%
  \BibitemOpen
  \bibfield  {author} {\bibinfo {author} {\bibfnamefont {H.~K.}\ \bibnamefont
  {Pal}}, \bibinfo {author} {\bibfnamefont {V.~I.}\ \bibnamefont {Yudson}}, \
  and\ \bibinfo {author} {\bibfnamefont {D.~L.}\ \bibnamefont {Maslov}},\
  }\bibfield  {title} {\enquote {\bibinfo {title} {{Resistivity of
  non-Galilean-invariant Fermi- and non-Fermi liquids}},}\ }\href@noop {}
  {\bibfield  {journal} {\bibinfo  {journal} {Lith. J. Phys.}\ }\textbf
  {\bibinfo {volume} {52}},\ \bibinfo {pages} {142} (\bibinfo {year}
  {2012}{\natexlab{a}})}\BibitemShut {NoStop}%
\bibitem [{\citenamefont {Sharma}\ \emph {et~al.}(2021)\citenamefont {Sharma},
  \citenamefont {Principi},\ and\ \citenamefont {Maslov}}]{Sharma:2021}%
  \BibitemOpen
  \bibfield  {author} {\bibinfo {author} {\bibfnamefont {Prachi}\ \bibnamefont
  {Sharma}}, \bibinfo {author} {\bibfnamefont {Alessandro}\ \bibnamefont
  {Principi}}, \ and\ \bibinfo {author} {\bibfnamefont {Dmitrii~L.}\
  \bibnamefont {Maslov}},\ }\bibfield  {title} {\enquote {\bibinfo {title}
  {{Optical conductivity of a Dirac-Fermi liquid}},}\ }\href {\doibase
  10.1103/PhysRevB.104.045142} {\bibfield  {journal} {\bibinfo  {journal}
  {Phys. Rev. B}\ }\textbf {\bibinfo {volume} {104}},\ \bibinfo {pages}
  {045142} (\bibinfo {year} {2021})}\BibitemShut {NoStop}%
\bibitem [{\citenamefont {Goyal}\ \emph {et~al.}(2023)\citenamefont {Goyal},
  \citenamefont {Sharma},\ and\ \citenamefont {Maslov}}]{Goyal:2023}%
  \BibitemOpen
  \bibfield  {author} {\bibinfo {author} {\bibfnamefont {Adamya~P.}\
  \bibnamefont {Goyal}}, \bibinfo {author} {\bibfnamefont {Prachi}\
  \bibnamefont {Sharma}}, \ and\ \bibinfo {author} {\bibfnamefont {Dmitrii~L.}\
  \bibnamefont {Maslov}},\ }\bibfield  {title} {\enquote {\bibinfo {title}
  {{Intrinsic optical absorption in Dirac metals}},}\ }\href {\doibase
  https://doi.org/10.1016/j.aop.2023.169355} {\bibfield  {journal} {\bibinfo
  {journal} {Annals of Physics}\ ,\ \bibinfo {pages} {169355}} (\bibinfo {year}
  {2023})}\BibitemShut {NoStop}%
\bibitem [{\citenamefont {Gurzhi}\ \emph {et~al.}(1982)\citenamefont {Gurzhi},
  \citenamefont {Kopeliovich},\ and\ \citenamefont {Rutkevich}}]{gurzhi:1982}%
  \BibitemOpen
  \bibfield  {author} {\bibinfo {author} {\bibfnamefont {R.N.}\ \bibnamefont
  {Gurzhi}}, \bibinfo {author} {\bibfnamefont {A.I.}\ \bibnamefont
  {Kopeliovich}}, \ and\ \bibinfo {author} {\bibfnamefont {S.~B.}\ \bibnamefont
  {Rutkevich}},\ }\bibfield  {title} {\enquote {\bibinfo {title} {Electric
  conductivity of two-dimensional metallic systems},}\ }\href
  {http://jetp.ac.ru/cgi-bin/e/index/e/56/1/p159?a=list} {\bibfield  {journal}
  {\bibinfo  {journal} {Sov. Phys.--JETP}\ }\textbf {\bibinfo {volume} {56}},\
  \bibinfo {pages} {159} (\bibinfo {year} {1982})}\BibitemShut {NoStop}%
\bibitem [{\citenamefont {{Gurzhi}}\ \emph {et~al.}(1987)\citenamefont
  {{Gurzhi}}, \citenamefont {{Kopeliovich}},\ and\ \citenamefont
  {{Rutkevich}}}]{gurzhi:1987}%
  \BibitemOpen
  \bibfield  {author} {\bibinfo {author} {\bibfnamefont {R.~N.}\ \bibnamefont
  {{Gurzhi}}}, \bibinfo {author} {\bibfnamefont {A.~I.}\ \bibnamefont
  {{Kopeliovich}}}, \ and\ \bibinfo {author} {\bibfnamefont {S.~B.}\
  \bibnamefont {{Rutkevich}}},\ }\bibfield  {title} {\enquote {\bibinfo {title}
  {{Kinetic properties of two-dimensional metal systems}},}\ }\href {\doibase
  10.1080/00018738700101002} {\bibfield  {journal} {\bibinfo  {journal} {Adv.
  Phys.}\ }\textbf {\bibinfo {volume} {36}},\ \bibinfo {pages} {221--270}
  (\bibinfo {year} {1987})}\BibitemShut {NoStop}%
\bibitem [{\citenamefont {Gurzhi}\ \emph {et~al.}(1995)\citenamefont {Gurzhi},
  \citenamefont {Kalinenko},\ and\ \citenamefont {Kopeliovich}}]{gurzhi:1995}%
  \BibitemOpen
  \bibfield  {author} {\bibinfo {author} {\bibfnamefont {R.~N.}\ \bibnamefont
  {Gurzhi}}, \bibinfo {author} {\bibfnamefont {A.~N.}\ \bibnamefont
  {Kalinenko}}, \ and\ \bibinfo {author} {\bibfnamefont {A.~I.}\ \bibnamefont
  {Kopeliovich}},\ }\bibfield  {title} {\enquote {\bibinfo {title}
  {Electron-electron momentum relaxation in a two-dimensional electron gas},}\
  }\href {\doibase 10.1103/PhysRevB.52.4744} {\bibfield  {journal} {\bibinfo
  {journal} {Phys. Rev. B}\ }\textbf {\bibinfo {volume} {52}},\ \bibinfo
  {pages} {4744--4747} (\bibinfo {year} {1995})}\BibitemShut {NoStop}%
\bibitem [{\citenamefont {Rosch}\ and\ \citenamefont
  {Howell}(2005)}]{rosch:2005}%
  \BibitemOpen
  \bibfield  {author} {\bibinfo {author} {\bibfnamefont {A.}~\bibnamefont
  {Rosch}}\ and\ \bibinfo {author} {\bibfnamefont {P.~C.}\ \bibnamefont
  {Howell}},\ }\bibfield  {title} {\enquote {\bibinfo {title} {Zero-temperature
  optical conductivity of ultraclean {F}ermi liquids and superconductors},}\
  }\href {\doibase 10.1103/PhysRevB.72.104510} {\bibfield  {journal} {\bibinfo
  {journal} {Phys. Rev. B}\ }\textbf {\bibinfo {volume} {72}},\ \bibinfo
  {pages} {104510} (\bibinfo {year} {2005})}\BibitemShut {NoStop}%
\bibitem [{\citenamefont {Rosch}(2006)}]{rosch:2006}%
  \BibitemOpen
  \bibfield  {author} {\bibinfo {author} {\bibfnamefont {A.}~\bibnamefont
  {Rosch}},\ }\bibfield  {title} {\enquote {\bibinfo {title} {Optical
  conductivity of clean metals},}\ }\href {\doibase 10.1002/andp.200510203}
  {\bibfield  {journal} {\bibinfo  {journal} {Annalen der Physik}\ }\textbf
  {\bibinfo {volume} {15}},\ \bibinfo {pages} {526--534} (\bibinfo {year}
  {2006})}\BibitemShut {NoStop}%
\bibitem [{\citenamefont {Briskot}\ \emph {et~al.}(2015)\citenamefont
  {Briskot}, \citenamefont {Sch\"utt}, \citenamefont {Gornyi}, \citenamefont
  {Titov}, \citenamefont {Narozhny},\ and\ \citenamefont
  {Mirlin}}]{briskot:2015}%
  \BibitemOpen
  \bibfield  {author} {\bibinfo {author} {\bibfnamefont {U.}~\bibnamefont
  {Briskot}}, \bibinfo {author} {\bibfnamefont {M.}~\bibnamefont {Sch\"utt}},
  \bibinfo {author} {\bibfnamefont {I.~V.}\ \bibnamefont {Gornyi}}, \bibinfo
  {author} {\bibfnamefont {M.}~\bibnamefont {Titov}}, \bibinfo {author}
  {\bibfnamefont {B.~N.}\ \bibnamefont {Narozhny}}, \ and\ \bibinfo {author}
  {\bibfnamefont {A.~D.}\ \bibnamefont {Mirlin}},\ }\bibfield  {title}
  {\enquote {\bibinfo {title} {Collision-dominated nonlinear hydrodynamics in
  graphene},}\ }\href {\doibase 10.1103/PhysRevB.92.115426} {\bibfield
  {journal} {\bibinfo  {journal} {Phys. Rev. B}\ }\textbf {\bibinfo {volume}
  {92}},\ \bibinfo {pages} {115426} (\bibinfo {year} {2015})}\BibitemShut
  {NoStop}%
\bibitem [{\citenamefont {Ledwith}\ \emph {et~al.}(2019)\citenamefont
  {Ledwith}, \citenamefont {Guo},\ and\ \citenamefont
  {Levitov}}]{levitov:2019}%
  \BibitemOpen
  \bibfield  {author} {\bibinfo {author} {\bibfnamefont {Patrick~J.}\
  \bibnamefont {Ledwith}}, \bibinfo {author} {\bibfnamefont {Haoyu}\
  \bibnamefont {Guo}}, \ and\ \bibinfo {author} {\bibfnamefont {Leonid}\
  \bibnamefont {Levitov}},\ }\bibfield  {title} {\enquote {\bibinfo {title}
  {The hierarchy of excitation lifetimes in two-dimensional {F}ermi gases},}\
  }\href {\doibase https://doi.org/10.1016/j.aop.2019.167913} {\bibfield
  {journal} {\bibinfo  {journal} {Ann. Phys.}\ }\textbf {\bibinfo {volume}
  {411}},\ \bibinfo {pages} {167913} (\bibinfo {year} {2019})}\BibitemShut
  {NoStop}%
\bibitem [{\citenamefont {Maslov}\ \emph {et~al.}(2011)\citenamefont {Maslov},
  \citenamefont {Yudson},\ and\ \citenamefont {Chubukov}}]{maslov:2011}%
  \BibitemOpen
  \bibfield  {author} {\bibinfo {author} {\bibfnamefont {Dmitrii~L.}\
  \bibnamefont {Maslov}}, \bibinfo {author} {\bibfnamefont {Vladimir~I.}\
  \bibnamefont {Yudson}}, \ and\ \bibinfo {author} {\bibfnamefont {Andrey~V.}\
  \bibnamefont {Chubukov}},\ }\bibfield  {title} {\enquote {\bibinfo {title}
  {{Resistivity of a Non-Galilean--Invariant Fermi Liquid near Pomeranchuk
  Quantum Criticality}},}\ }\href {\doibase 10.1103/PhysRevLett.106.106403}
  {\bibfield  {journal} {\bibinfo  {journal} {Phys. Rev. Lett.}\ }\textbf
  {\bibinfo {volume} {106}},\ \bibinfo {pages} {106403} (\bibinfo {year}
  {2011})}\BibitemShut {NoStop}%
\bibitem [{\citenamefont {Guo}\ \emph {et~al.}(2022)\citenamefont {Guo},
  \citenamefont {Patel}, \citenamefont {Esterlis},\ and\ \citenamefont
  {Sachdev}}]{Guo:2022}%
  \BibitemOpen
  \bibfield  {author} {\bibinfo {author} {\bibfnamefont {Haoyu}\ \bibnamefont
  {Guo}}, \bibinfo {author} {\bibfnamefont {Aavishkar~A.}\ \bibnamefont
  {Patel}}, \bibinfo {author} {\bibfnamefont {Ilya}\ \bibnamefont {Esterlis}},
  \ and\ \bibinfo {author} {\bibfnamefont {Subir}\ \bibnamefont {Sachdev}},\
  }\bibfield  {title} {\enquote {\bibinfo {title} {Large-${N}$ theory of
  critical fermi surfaces. ii. {C}onductivity},}\ }\href {\doibase
  10.1103/PhysRevB.106.115151} {\bibfield  {journal} {\bibinfo  {journal}
  {Phys. Rev. B}\ }\textbf {\bibinfo {volume} {106}},\ \bibinfo {pages}
  {115151} (\bibinfo {year} {2022})}\BibitemShut {NoStop}%
\bibitem [{\citenamefont {Shi}\ \emph {et~al.}(2022)\citenamefont {Shi},
  \citenamefont {Goldman}, \citenamefont {Else},\ and\ \citenamefont
  {Senthil}}]{Senthil:2022}%
  \BibitemOpen
  \bibfield  {author} {\bibinfo {author} {\bibfnamefont {Zhengyan~Darius}\
  \bibnamefont {Shi}}, \bibinfo {author} {\bibfnamefont {Hart}\ \bibnamefont
  {Goldman}}, \bibinfo {author} {\bibfnamefont {Dominic~V.}\ \bibnamefont
  {Else}}, \ and\ \bibinfo {author} {\bibfnamefont {T.}~\bibnamefont
  {Senthil}},\ }\bibfield  {title} {\enquote {\bibinfo {title} {{Gifts from
  anomalies: Exact results for Landau phase transitions in metals}},}\ }\href
  {\doibase 10.21468/SciPostPhys.13.5.102} {\bibfield  {journal} {\bibinfo
  {journal} {SciPost Phys.}\ }\textbf {\bibinfo {volume} {13}},\ \bibinfo
  {pages} {102} (\bibinfo {year} {2022})}\BibitemShut {NoStop}%
\bibitem [{\citenamefont {Hertz}(1976)}]{hertz:1976}%
  \BibitemOpen
  \bibfield  {author} {\bibinfo {author} {\bibfnamefont {John~A.}\ \bibnamefont
  {Hertz}},\ }\bibfield  {title} {\enquote {\bibinfo {title} {Quantum critical
  phenomena},}\ }\href {\doibase 10.1103/PhysRevB.14.1165} {\bibfield
  {journal} {\bibinfo  {journal} {Phys. Rev. B}\ }\textbf {\bibinfo {volume}
  {14}},\ \bibinfo {pages} {1165--1184} (\bibinfo {year} {1976})}\BibitemShut
  {NoStop}%
\bibitem [{\citenamefont {{Millis}}(1993)}]{millis:1993}%
  \BibitemOpen
  \bibfield  {author} {\bibinfo {author} {\bibfnamefont {A.~J.}\ \bibnamefont
  {{Millis}}},\ }\bibfield  {title} {\enquote {\bibinfo {title} {{Effect of a
  nonzero temperature on quantum critical points in itinerant fermion
  systems}},}\ }\href {\doibase 10.1103/PhysRevB.48.7183} {\bibfield  {journal}
  {\bibinfo  {journal} {\prb}\ }\textbf {\bibinfo {volume} {48}},\ \bibinfo
  {pages} {7183--7196} (\bibinfo {year} {1993})}\BibitemShut {NoStop}%
\bibitem [{\citenamefont {Kim}\ \emph {et~al.}(1994)\citenamefont {Kim},
  \citenamefont {Furusaki}, \citenamefont {Wen},\ and\ \citenamefont
  {Lee}}]{kim:1994}%
  \BibitemOpen
  \bibfield  {author} {\bibinfo {author} {\bibfnamefont {Yong~Baek}\
  \bibnamefont {Kim}}, \bibinfo {author} {\bibfnamefont {Akira}\ \bibnamefont
  {Furusaki}}, \bibinfo {author} {\bibfnamefont {Xiao-Gang}\ \bibnamefont
  {Wen}}, \ and\ \bibinfo {author} {\bibfnamefont {Patrick~A.}\ \bibnamefont
  {Lee}},\ }\bibfield  {title} {\enquote {\bibinfo {title} {Gauge-invariant
  response functions of fermions coupled to a gauge field},}\ }\href {\doibase
  10.1103/PhysRevB.50.17917} {\bibfield  {journal} {\bibinfo  {journal} {Phys.
  Rev. B}\ }\textbf {\bibinfo {volume} {50}},\ \bibinfo {pages} {17917--17932}
  (\bibinfo {year} {1994})}\BibitemShut {NoStop}%
\bibitem [{\citenamefont {Chubukov}\ and\ \citenamefont
  {Maslov}(2017)}]{chubukov:2017}%
  \BibitemOpen
  \bibfield  {author} {\bibinfo {author} {\bibfnamefont {Andrey~V.}\
  \bibnamefont {Chubukov}}\ and\ \bibinfo {author} {\bibfnamefont {Dmitrii~L.}\
  \bibnamefont {Maslov}},\ }\bibfield  {title} {\enquote {\bibinfo {title}
  {Optical conductivity of a two-dimensional metal near a quantum critical
  point: The status of the extended {D}rude formula},}\ }\href {\doibase
  10.1103/PhysRevB.96.205136} {\bibfield  {journal} {\bibinfo  {journal} {Phys.
  Rev. B}\ }\textbf {\bibinfo {volume} {96}},\ \bibinfo {pages} {205136}
  (\bibinfo {year} {2017})}\BibitemShut {NoStop}%
\bibitem [{\citenamefont {Shi}\ \emph {et~al.}(2023)\citenamefont {Shi},
  \citenamefont {Else}, \citenamefont {Goldman},\ and\ \citenamefont
  {Senthil}}]{Senthil:2023}%
  \BibitemOpen
  \bibfield  {author} {\bibinfo {author} {\bibfnamefont {Zhengyan~Darius}\
  \bibnamefont {Shi}}, \bibinfo {author} {\bibfnamefont {Dominic~V.}\
  \bibnamefont {Else}}, \bibinfo {author} {\bibfnamefont {Hart}\ \bibnamefont
  {Goldman}}, \ and\ \bibinfo {author} {\bibfnamefont {T.}~\bibnamefont
  {Senthil}},\ }\bibfield  {title} {\enquote {\bibinfo {title} {{Loop current
  fluctuations and quantum critical transport}},}\ }\href {\doibase
  10.21468/SciPostPhys.14.5.113} {\bibfield  {journal} {\bibinfo  {journal}
  {SciPost Phys.}\ }\textbf {\bibinfo {volume} {14}},\ \bibinfo {pages} {113}
  (\bibinfo {year} {2023})}\BibitemShut {NoStop}%
\bibitem [{\citenamefont {Pal}\ \emph {et~al.}(2012{\natexlab{b}})\citenamefont
  {Pal}, \citenamefont {Yudson},\ and\ \citenamefont {Maslov}}]{pal:2012}%
  \BibitemOpen
  \bibfield  {author} {\bibinfo {author} {\bibfnamefont {H.~K.}\ \bibnamefont
  {Pal}}, \bibinfo {author} {\bibfnamefont {V.~I.}\ \bibnamefont {Yudson}}, \
  and\ \bibinfo {author} {\bibfnamefont {D.~L.}\ \bibnamefont {Maslov}},\
  }\bibfield  {title} {\enquote {\bibinfo {title} {{Effect of electron-electron
  interaction on surface transport in the Bi${}_{2}$Te${}_{3}$ family of
  three-dimensional topological insulators}},}\ }\href {\doibase
  10.1103/PhysRevB.85.085439} {\bibfield  {journal} {\bibinfo  {journal} {Phys.
  Rev. B}\ }\textbf {\bibinfo {volume} {85}},\ \bibinfo {pages} {085439}
  (\bibinfo {year} {2012}{\natexlab{b}})}\BibitemShut {NoStop}%
\bibitem [{\citenamefont {Pimenov}\ \emph {et~al.}(2022)\citenamefont
  {Pimenov}, \citenamefont {Kamenev},\ and\ \citenamefont
  {Chubukov}}]{Pimenov:2022}%
  \BibitemOpen
  \bibfield  {author} {\bibinfo {author} {\bibfnamefont {Dimitri}\ \bibnamefont
  {Pimenov}}, \bibinfo {author} {\bibfnamefont {Alex}\ \bibnamefont {Kamenev}},
  \ and\ \bibinfo {author} {\bibfnamefont {Andrey~V.}\ \bibnamefont
  {Chubukov}},\ }\bibfield  {title} {\enquote {\bibinfo {title} {Quasiparticle
  scattering in a superconductor near a nematic critical point: Resonance mode
  and multiple attractive channels},}\ }\href {\doibase
  10.1103/PhysRevLett.128.017001} {\bibfield  {journal} {\bibinfo  {journal}
  {Phys. Rev. Lett.}\ }\textbf {\bibinfo {volume} {128}},\ \bibinfo {pages}
  {017001} (\bibinfo {year} {2022})}\BibitemShut {NoStop}%
\end{thebibliography}%

\end{document}